\def\Granadadep{Departamento de F\'\i sica Te\'orica y del Cosmos, Facultad
de Ciencias, Universidad de Granada, 
Campus de Fuentenueva, Granada 18002, Spain. E-mail: calixto@ugr.es}

\def\Granadainst{Instituto ``Carlos I" de F\'\i sica 
Te\'orica y Computacional, Facultad
de Ciencias, Universidad de Granada, 
Campus de Fuentenueva, Granada 18002, Spain. http://www.ugr.es/\~{}carlos1}

\def\Valencia{IFIC, Centro Mixto Universidad de Valencia-CSIC, Burjasot
              46100-Valencia, Spain. E-mail: valdaya@ugr.es}

\def\Comision{Work partially supported by the DGICYT under contracts PB92-1055,
PB92-0302, PB95-1201 and PB95-0947}
\def\Salamanca{Facultad de F\'\i sicas, edificio Triling\"ue, Universidad de 
Salamanca, 37008 Salamanca, Spain. E-mail: cervero@rs6000.usal.es}

\def\cp{{{}_{\cal P}}}
\def\be{\begin{equation}}
\def\ee{\end{equation}}
\def\bea{\begin{eqnarray}}
\def\eea{\end{eqnarray}}
\def\p{\partial}
\def\TG{\tilde{G}}
\def\tg{\tilde{g}}
\def\th{\tilde{h}}
\def\hs{\bar{s}}

\def\TQR{\tilde{Q}^R}

\def\hz{\bar{z}}
\def\hzc{{\bar{z}^*}}
\def\hn{{\bar{n}}}
\def\hm{{\bar{m}}}

\def\f{\frac}
\def\l[{\left[}
\def\r]{\right]}
\def\XL{X^L}
\def\XR{X^R}
\def\TPIOL{\tilde{\Pi}_0^L}
\def\TPIL{\tilde{\Pi}_1^L}
\def\TX1L{\tilde{X}_1^L}
\def\TXOL{\tilde{X}_0^L}
\def\TXR{\tilde{X}^R}
\def\TXL{\tilde{X}^L}
\def\nn{\nonumber}
\def\ni{\noindent}
\def\heta{\bar{\eta}}
\def\hc{\bar{\alpha}}
\def\hcc{{\bar{\alpha}^*}}
\def\DL{D^L}
\def\ML{M^L}
\def\POL{P_0^L}
\def\PL{P_1^L}
\def\KOL{K_0^L}
\def\KL{K_1^L}
\def\TDL{\tilde{D}^L}
\def\TML{\tilde{M}^L}
\def\TPOL{\tilde{P}_0^L}
\def\TPL{\tilde{P}_1^L}
\def\TKOL{\tilde{K}_0^L}
\def\TKL{\tilde{K}_1^L}
\def\TDR{\tilde{D}^R}
\def\TMR{\tilde{M}^R}
\def\TPOR{\tilde{P}_0^R}
\def\TPR{\tilde{P}_1^R}
\def\TKOR{\tilde{K}_0^R}
\def\TKR{\tilde{K}_1^R}
\def\TDL{\tilde{D}^L}
\def\TML{\tilde{M}^L}
\def\TPOL{\tilde{P}_0^L}
\def\TPL{\tilde{P}_1^L}
\def\TKOL{\tilde{K}_0^L}
\def\TKL{\tilde{K}_1^L}
\def\TDLS{\tilde{D}^{L(2)}}
\def\TMLS{\tilde{M}^{L(2)}}
\def\TPOLS{\tilde{P}_0^{L(2)}}
\def\TPLS{\tilde{P}_1^{L(2)}}
\def\TKOLS{\tilde{K}_0^{L(2)}}
\def\TKLS{\tilde{K}_1^{L(2)}}
\def\TDRS{\tilde{D}^{R(2)}}
\def\TMRS{\tilde{M}^{R(2)}}
\def\TPORS{\tilde{P}_0^{R(2)}}
\def\TPRS{\tilde{P}_1^{R(2)}}
\def\TKORS{\tilde{K}_0^{R(2)}}
\def\TKRS{\tilde{K}_1^{R(2)}}
\def\TXLS{\tilde{X}^{L(2)}}
\def\TXRS{\tilde{X}^{R(2)}}
\def\A{\rho^{(N)}_{mn;\hm\hn}}
\def\Ac{\rho^{(N)*}_{mn;\hm\hn}}
\def\a{{\hat{a}}}
\def\ac{{\hat{a}}^{\dag}}
\def\cc{\alpha^*}
\def\c{\alpha}
\def\hmu{\bar{\mu}}
\def\z{z}
\def\ere{r}
\def\vs{\varsigma}
\documentstyle[11pt,epsf]{article}

\begin{document}

\begin{titlepage}

%\begin{flushright}
%{\it  hep-th/9705049}
%\end{flushright}
\vskip 1cm
%\vfill

\begin{center}
  
{\bf VACUUM RADIATION 
AND SYMMETRY BREAKING IN CONFORMALLY INVARIANT 
QUANTUM FIELD THEORY $^1$ }
\end{center}

\bigskip
\bigskip

\centerline{V. Aldaya$^{2,3}$,\,\,  M. Calixto$^{2,4}$,\, 
and J.M. Cerver\'o$^5$}

\bigskip
%\centerline{June 1995}
\bigskip

\bigskip

\begin{center}
{\bf Abstract}
\end{center}
The underlying reasons for the difficulty of unitarily 
implementing the whole conformal group $SO(4,2)$ in a massless {\it Quantum}  
Field Theory (QFT) on the Minkowski space are investigated in this paper. 
Firstly, we demonstrate that the 
singular action of the subgroup of special conformal transformations (SCT), 
on the standard Minkowski space $M$, cannot be 
primarily associated with the vacuum 
radiation problems, the reason being more 
profound and related to the dynamical breakdown of part of the conformal 
symmetry (the SCT subgroup, to be more precise) when 
representations of null mass are selected inside the representations of the 
whole conformal group. Then we show how the vacuum of the massless QFT 
{\it radiates} under the action of SCT (usually interpreted as transitions 
to a uniformly accelerated frame) and we calculate exactly the spectrum of the 
outgoing particles, which proves to be a generalization of the Planckian one, 
this recovered as a given limit.

\vskip 0.5cm
\ni PAC numbers: 03.65.Fd, 03.70.+k\\
\ni Keywords: vacuum radiation, conformal symmetry  breaking, uniformly 
accelerated frames, group quantization. 
\footnotetext[1]{\Comision}
\footnotetext[2]{\Granadainst} \footnotetext[3]{\Valencia}
\footnotetext[4]{\Granadadep}\footnotetext[5]{\Salamanca}

\vfil

\end{titlepage}

\vfil\pagebreak

\section{Introduction}

The conformal group $SO(4,2)$ has ever been recognized as a symmetry of the 
Maxwell equations for {\it classical} electro-dynamics 
\cite{Maxwell}, and more recently considered as 
an invariance of general, non-abelian, maseless gauge field theories at 
the classical 
level. However, the {\it quantum} theory raises, in general, 
serious problems in 
the implementation of conformal symmetry, and much work has been devoted to 
the study of the physical reasons for that (see e.g. Ref. \cite{Fronsdal}). 
Basically, the main trouble 
associated with this quantum symmetry (at the second quantization level) 
lies in the difficulty of finding 
a vacuum,  which is {\it stable} under special conformal 
transformations acting on 
the Minkowski space in the form:
\be
x^\mu \rightarrow {x'}^\mu=\f{x^\mu+c^\mu
x^2}{\sigma(x,c)},\;\;\;\sigma(x,c)=1+2c x+c^2 x^2
\label{confact}.
\ee
\ni These transformations, which can be interpreted as transitions to systems 
of relativistic, uniformly accelerated observers (see e.g. Ref. \cite{Hill}), 
cause {\it vacuum radiation}, a phenomenon 
analogous to the Fulling-Unruh effect
\cite{Fulling,Unruh} in a non-inertial reference  frame. To be more precise, 
if $a(k),a^+(k)$ are the Fourier components of a scalar massless field 
$\phi(x)$, satisfying the equation
\be
\eta^{\mu\nu}\p_{\mu}\p_\nu \phi(x)=0 \, ,\label{kg}
\ee
then, the Fourier components $a'(k), {a'}^+(k)$ of the transformed field 
$\phi'(x')=\sigma^{-l}(x,c)\phi(x)$ by (\ref{confact}) ($l$ being the
conformal dimension) are expressed in terms of both $a(k),a^+(k)$ through a 
Bogolyubov transformation
\be
a'(\lambda)=\int{dk\l[ A_\lambda(k)a(k)+B_\lambda(k)a^+(k)\r]}\,. \label{bog}
\ee
\ni In the second quantized theory,  the vacuum states defined by the
conditions $a(k)|0\rangle   =0$ and $a'(\lambda)|0'\rangle   =0$, 
are not identical if the coefficients $B_\lambda(k)$ in (\ref{bog})
are not zero. In this case the new vacuum has a non-trivial content of 
untransformed particle states.
 
This situation is always present when quantizing field theories in
curved space as well as in flat space, whenever some kind of global mutilation
of the space is involved. This is the case of the 
natural quantization in Rindler
coordinates \cite{Birrell}, which leads to a quantization inequivalent to the
normal Minkowski quantization, or that of a quantum field in a box, where a
dilatation produces a rearrangement of the vacuum \cite{Fulling}. Nevertheless,
it must be stressed that the situation for SCT is more peculiar. 
The rearrangement of the vacuum in a massless
QFT due to SCT, even though they are a 
symmetry of the classical system, behaves as if the conformal group were 
{\it spontaneously broken}, and this fact can be 
interpreted as a kind of topological 
{\it anomaly}.

Thinking of the underlying reasons for this anomaly, we are tempted to make
the singular action of the transformations (\ref{confact}) in Minkowski space
responsible for it, as has been in fact pointed out in \cite{rusos}. However, a
deeper analysis of the interconnection between symmetry and quantization will 
reveal a more profound obstruction to the 
possibility of implementing unitarily 
STC  in a generalized Minkowski space, free 
from singularities, when conformally invariant fields are forced to evolve in 
time. This way, the quantum time evolution  
itself destroys the conformal symmetry, 
leading to some sort of {\it dynamical symmetry breaking} which 
preserves the Weyl 
subgroup (Poincar\'e + dilatations).

This obstruction is traced back to the 
impossibility of representing the entire 
$SO(4,2)$ group unitarily and irreducibly on a space of functions depending 
arbitrarily on $\vec{x}$ (see e.g.  Ref. \cite{Fronsdal}), so that a Cauchy 
surface determines the evolution in 
time. Natural representations, however, can be constructed by means of wave 
functions having support on the hole space-time and evolving in some kind of 
{\it proper time}. 

From the point of view of particle quantum mechanics 
(or ``first" quantization),
the free arguments of wave functions in the 
configuration-space ``representation"
correspond to half of the canonically conjugated variables in phase space 
(or classical solution manifold), and this phase space is usually defined as a 
co-adjoint orbit of the basic symmetry 
group characterizing the physical system.
Thus, for instance, for the Galilei or 
Poincar\'e group the phase space associated 
with massive spinless particles has dimension $6$ and the corresponding wave 
functions in configuration space have the time variable factorized out. 
However, as mentioned above, this is not the case for the conformal group, for 
which we shall realize that {\it time is a quantum observable} subject to 
uncertainty relations; this fact extends covariance rules to the quantum 
domain.

The present study is developed in the 
framework of a Group Approach to Quantization
(GAQ)\cite{GAQ,Ramirez}, which proves 
to be specially suitable for facing those quantization problems
arising from specific symmetry requirements. 
Furthermore, this formalism has the virtue
of providing in a natural way the space on 
which the wave functions are defined.
A very brief report on GAQ is presented in Sec. 2. In Sec. 3 we apply this 
quantization technique to the particular case of the group 
$SO(2,2)$, which is the 1+1 dimensional version of 
the $SO(4,2)$ symmetry. Although the conformal symmetry in 1+1 dimensions is 
far richer, we proceed in a way that can be straightforwardly extended  to the 
realistic dimension. In this example we show how a (compact) configuration 
space, locally isomorphic to Minkowski space time, can be found inside the 
$SO(2,2)$ group manifold on which the whole conformal group
acts without singularities. We also prove 
that the unitarity of the irreducible 
representations of  $SO(2,2)$ requires the dynamical character of the time 
variable, or that which is similar, prevents the existence of a conformally 
invariant quantum evolution equation in the time variable. We examine two 
cases corresponding to non-compact and compact ``proper time" dynamics in 
Subsec. 3.1 and 3.2 , respectively. Sec. 4 is devoted to the application of 
GAQ to a very special infinite-dimensional Lie group  
$\tilde{G}^{(2)}({\cal H}(\TG),\TG)$ directly attached to the quantum 
mechanical Hilbert space ${\cal H}(\TG)$ of a ``first"-quantized system 
characterized by the {\it quantizing group} $\TG$ (a central extension of 
$G=SO(2,2)$, for the present case). This mechanism is nothing other than a 
group version of the ``second"-quantization algorithm. 
With this algorithm at hand we formulate in Sec. 4.1 a conformally invariant 
quantum field theory and, in Sec. 4.2, we investigate the effect of a SCT on 
a Weyl vacuum and the associated radiation phenomenon. 
We calculate exactly the spectrum of  an accelerated 
Weyl vacuum, which proves to be a generalization of 
the black body spectrum, this recovered as a given limit.
Final comments are presented in the last Sec. 5.

\section{Quantization on a group $\TG$}

The starting point of GAQ is a group $\TG$ (the quantizing group) 
with a principal fibre bundle 
structure $\TG(B,T)$, having $T$ as the structure group and $B$ being 
the base. The group $T$ generalizes 
the phase invariance of Quantum Mechanics. Although the situation can be more 
general \cite{Ramirez}, we shall
start with the rather general case in which $\TG$ is a central extension of
a group $G$ by $T$ [$T=U(1)$ or even $T=C^*=\Re^+\otimes U(1)$]. 
For the one-parametric group 
$T=U(1)$, the group law for $\TG=\{\tilde{g}=(g,\zeta)/ g\in G, 
\zeta\in U(1)\}$ adopts the following 
form:
\be
\tilde{g}'*\tilde{g}=(g'*g,\zeta'\zeta e^{i\xi(g',g)})
\ee
\ni where $g''=g'*g$ is the group operation in $G$ and $\xi(g',g)$ is a 
two-cocycle of $G$ on $\Re$ fulfilling:
\be
\xi(g_2,g_1)+\xi(g_2*g_1,g_3)=\xi(g_2,g_1*g_3)+\xi(g_1,g_3)\;\;, g_i\in G. 
\ee
\ni In the general theory of central extensions \cite{Extensiones}, two 
two-cocycles are said to be equivalent if they differ in a coboundary, i.e. a 
cocycle which can be written in the form  $\xi(g',g)=\delta(g'*g)-
\delta(g')-\delta(g)$, where $\delta(g)$ is 
called the generating function of the coboundary. However, although cocycles 
differing on a coboundary lead to equivalent central extensions as such, 
there are some coboundaries which provide a non-trivial connection on the 
fibre bundle $\TG$ and Lie-algebra structure constants different from that 
of the direct product $G\otimes U(1)$. These are generated by a function 
$\delta$ with a non-trivial gradient at the 
identity, and can be divided into equivalence Pseudo-cohomology subclasses: 
two pseudo-cocycles are equivalent if they differ in a coboundary generated 
by a function with trivial gradient at the identity \cite{Saletan,Pseudoco,
Marmo}. Pseudo-cohomology plays an important role in the theory of 
finite-dimensional semi-simple group, as they have trivial cohomology. For 
them, Pseudo-cohomology classes are associated with coadjoint orbits 
\cite{Marmo}.

The right and left finite actions of the group $\TG$ 
on itself provide two sets 
of mutually commuting (left- and right-, respectively) invariant vector fields:
\be
\TXL_{\tg^i}=\left.\f{\p {\tg''}{}^j}{\p \tg^i}\right|_{\tg=e}
\f{\p}{\p \tg^j},\;\;\;\TXR_{\tg^i}=
\left.\f{\p {\tg''}{}^j}{\p {\tg'}{}^i}\right|_{\tg'=e}\f{\p}{\p \tg^j},
\;\;\; \l[\TXL_{\tg^i},\TXR_{\tg^j}\r]=0,\label{txlr1}
\ee
\ni where $\{\tg^j\}$ is a parameterization of $\TG$.
The GAQ program continues finding the left-invariant 1-form $\Theta$ (the 
{\it Quantization 1-form})
associated with the central generator $\TXL_\zeta=\TXR_\zeta, \zeta\in T$, 
i.e. the $T$-component
$\tilde{\theta}^{L(\zeta)}$ of the canonical left-invariant 1-form
$\tilde{\theta}^L$ on $\tilde{G}$. This constitutes the generalization of the 
Poincar\'e-Cartan form of Classical Mechanics (see \cite{Abraham}).
The differential $d\Theta$ is a
{\it presymplectic} form and its {\it characteristic module}, $Ker\Theta\cap
Ker {}d\Theta$, is generated by a left subalgebra ${\cal G}_\Theta$ named {\it
characteristic subalgebra}. The quotient $(\TG, \Theta)/{\cal G}_\Theta$ is a
{\it quantum manifold} in the sense of Geometric Quantization 
\cite{GQ}. The trajectories generated by the vector fields  in 
${\cal G}_\Theta$ constitute the generalized equations of motion of 
the theory (temporal evolution, rotations, etc...), and the ``Noether" 
invariants under those equations are $F_{\tg^j}\equiv i_{\TXR_{\tg^j}}\Theta$, 
i.e. the contraction of right-invariant vector fields with 
the Quantization 1-form. 

Let ${\cal B}(\TG)$ be the set of complex-valued
$T$-{\it functions} on $\TG$ in the sense
of principal bundle theory: 
\be
\psi(\zeta*\tg)=D_{T}(\zeta)\psi(\tg), \;\;\zeta\in T\,,\label{tcondition}
\ee
\ni where $D_{T}$ is the natural representation of $T$ on the 
complex numbers $C$. 
The representation of $\TG$ on ${\cal B}(\TG)$ generated
by ${\tilde{\cal G}}^R=\{\TXR\}$ is called {\it Bohr Quantization} and is 
{\it reducible}. The reduction can be achieved by means of the 
restrictions imposed by a {\it full polarization} ${\cal P}$:
\be
\TXL\psi_\cp=0,\;\;\forall \TXL\in {\cal P}\,,\label{defpol}
\ee
\ni which is a maximal, horizontal (excluding $\TXL_\zeta$) left
subalgebra of ${\tilde{\cal G}}^L$ which contains ${\cal G}_\Theta$. 
It should be noted that the existence of a full polarization, containing 
the whole subalgebra ${\cal G}_\Theta$, is not guaranteed. In case of such a
breakdown, called {\it anomaly}, or simply for the desire of 
choosing of a preferred representation space, a higher-order polarization 
has to be imposed \cite{HO}. A higher-order polarization 
is a maximal, horizontal subalgebra of the left enveloping algebra 
$U\tilde{{\cal G}}^L$ which contains ${\cal G}_\Theta$. 
   
The group $\TG$ is irreducibly represented on the space 
${\cal H}(\TG)\equiv\left\{|\psi\rangle \right\}$ of polarized wave 
functions, and on its dual  
${\cal H}^*(\TG)\equiv\left\{\langle \psi|\right\}$. If we denote by 
\be
\langle \tg_\cp|\psi\rangle\equiv\psi_\cp(\tg) \,,\,\,
\langle \psi'|\tg_\cp\rangle\equiv\psi'{}^*_\cp(\tg)
\ee
\ni the coordinates of the ``ket" $|\psi\rangle $ and the ``bra" 
$\langle \psi'|$ in a representation defined through 
a polarization ${\cal P}$ (first- or higher-order), then, a scalar product 
on ${\cal H}(\TG)$ can be naturally defined as:

\be
\langle \psi'|\psi\rangle \equiv
\int_{\TG}{ v(\tg)\psi_\cp'{}^*(\tg)\psi_\cp(\tg)},
\ee
\ni where  
\be
 v(\tg)\equiv\theta^L_{\tg^i}\wedge\stackrel{\hbox{dim}(\TG)}{...}
\wedge\theta^L_{\tg^n}\label{volumen}
\ee
\ni is the left-invariant integration volume in $\TG$ and
\be
1=\int_{\TG}{|\tg_\cp\rangle  v(\tg)\langle \tg_\cp|}\label{closure}
\ee
\ni formally represents a {\it closure} relation.
A direct computation proves that, with this scalar product, 
the group $\TG$ is unitarily represented through the 
{\it left} finite action ($\rho$ denotes the representation)
\be
\langle \tg_\cp|\rho(\tg')|\psi\rangle \equiv\psi_\cp(\tg'{}^{-1}*\tg)\,. 
\label{leftaction}
\ee
\ni The {\it adjoint} action is then defined as 
\be
\langle \psi'|\rho^{\dag} (\tg')|\psi\rangle\equiv
\langle \psi|\rho(\tg')|\psi'\rangle^* ,\;\;\hbox{i.e}\;\;
\langle \tg_\cp|\rho^{\dag} (\tg')|\psi\rangle =\psi_\cp(\tg'*\tg)
\,.\label{adjoint}
\ee

We can relate the coordinates of $|\psi\rangle $ in two 
given representations, corresponding with two different polarizations 
${\cal P}_1$ and ${\cal P}_2$, as follows:
\be
\psi_{\cp_1}(\tg)=\langle \tg_{\cp_1}|\psi\rangle =
\int_{\TG}{ v(\tg')\langle \tg_{\cp_1}|\tg_{\cp_2}'\rangle 
\langle \tg_{\cp_2}'|\psi\rangle }\equiv\int_{\TG}{ v(\tg')
{{\Delta}}_{\cp_1\cp_2}(\tg,\tg')\psi_{\cp_2}(\tg')}\,,\label{polchange}
\ee
\ni where ${{\Delta}}_{\cp_1\cp_2}(\tg,\tg')$ is a 
``polarization changing" operator.
An explicit expression of ${{\Delta}}_{\cp_1\cp_2}$ 
can be obtained by making use of a basis 
$\left\{|n\rangle \right\}_{n\in I}$ ($I$ is a set of indices) of 
${\cal H}(\TG)$, as follows:
\be
{{\Delta}}_{\cp_1\cp_2}(\tg,\tg')=\langle \tg_{\cp_1}|\tg_{\cp_2}'\rangle =
\sum_{n\in I}\psi_{\cp_1,n}^*(\tg)\psi_{\cp_2,n}(\tg')\,,\label{operchan}
\ee
\ni where $\psi_{\cp_i,n}(\tg)\equiv\langle \tg_{\cp_i}|n\rangle $ are 
the coordinates of $|n\rangle $ in a polarization ${\cal P}_i$.

\vskip 0.2cm
{\it Constraints} are consistently incorporated into the theory by 
enlarging  the structure group $T$ (which always includes $U(1)$), i.e. 
through  $T$-function conditions:
\be
\rho(\tilde{t})|\psi\rangle =D^{(\epsilon)}_T(\tilde{t})|\psi\rangle\,,\;\;
\tilde{t}\in T 
\ee
\ni or, for continuous transformations,
\be
\TXR_{\tilde{t}}|\psi\rangle =dD^{(\epsilon)}_T(\tilde{t})|\psi\rangle \;, 
\label{const}
\ee 
$D^{(\epsilon)}_T$ means a specific representation of $T$  [the index 
$\epsilon$ parametrizes different (inequivalent) quantizations] and 
$dD^{(\epsilon)}_T$ is its differential.

 It is obvious that, for a non-central structure 
group $T$, not all the right operators $\TXR_{\tg}$ will preserve 
these constraints; a sufficient condition for a subgroup  
$\TG_T\subset\TG$ to preserve the constraints is (see \cite{FracHall}):
\be
 \l[\TG_T,T\r]\subset \hbox{Ker}D^{(\epsilon)}_T
\ee
\ni  [note that, for the trivial representation of $T$, 
the subgroup $\TG_T$ is nothing other than the {\it normalizer} of $T$]. 
$\TG_T$ takes part of the set of {\it good} operators \cite{Ramirez}, of
the enveloping algebra $U\tilde{{\cal G}}^R$ in general, for which 
the subgroup $T$ behaves as a {\it gauge} group (see \cite{config2} for a 
thorough study of gauge symmetries and constraints from the point of 
view of GAQ). A more general situation can be posed when the 
constraints are lifted to the higher-order level, not necessarily first 
order as in (\ref{const}), that is, they are a subalgebra of the 
right enveloping 
algebra $U\tilde{{\cal G}}^R$. An interesting example of this last case 
arises when one selects representations labelled by a value $\epsilon$ 
of some Casimir operator $Q$ of a subgroup $\TG_Q$ of $\TG$. 
This is exactly the case that interests us: null mass representations 
($\epsilon=m=0$) of the Poincar\'e group 
($\TG_Q=SO(3,1)\otimes_s T_4\,,\;Q=P_\nu P^\nu$) inside 
the conformal group ($\TG=SO(4,2)$).

In the more general case in which $T$ is not a trivial central extension,
$T \neq \check{T} \times U(1)$, where $\check{T} \equiv T/U(1)$ -i.e. $T$ 
contains second-class constraints- the conditions (\ref{const}) are not all
compatible and we must select a subgroup $T_B = T_p \times U(1)$, where
$T_p$ is the subgroup associated with a right polarization subalgebra of the
central extension $T$ (see \cite{Ramirez}).
   
For simplicity, we have sometimes made use of infinitesimal (geometrical) 
concepts, but all this language can be translated to their finite (algebraic) 
counterparts (see \cite{Ramirez}), a desirable way of 
proceeding when discrete  
transformations are incorporated into the theory.

\section{Conformally invariant Quantum Mechanics}

Conformally invariant Quantum Mechanics (in 1+1D) will be developed by 
finding the unitary irreducible representations of the centrally extended
$SO(2,2)$ group in exactly the same way the Hilbert space of the Galilean 
particle is obtained from the unitary irreducible representations of the
centrally extended Galilei group (see e.g. Ref. \cite{GAQ}). 
The configuration space of the theory or, rather, an 
analytic continuation of Minkowski space, will arise as a homogeneous 
space of the group, on which the wave functions supporting the irreducible 
representation take arguments.

Except for discrete symmetries, which are not 
relevant at the Lie algebra level,
$SO(2,2)\sim SU(1,1)\otimes SU(1,1)$ so that 
we shall look at the structure of
\be
SU(1,1)=\left\{ U= \left( \begin{array}{cc} z_1&z_2\\z_2^*&z_1^*\end{array}
\right) ,z_i,z_i^* \in C/ \det(U)=|z_1|^2-|z_2|^2=1\right\} \,.
\ee
\ni $SU(1,1)$ is a fibre bundle with fibre $U(1)$ and base the hyperboloid. 
A system of coordinates adapted to this fibration is the following:
\be
\eta\equiv\f{z_1}{|z_1|}, \;\; \c\equiv\f{z_2}{z_1},\;\;
\c^*\equiv\f{z_2^*}{z_1^*},\;\;\;\; \eta\in U(1),\;\;\c,\c^*\in D_1,
\ee
\ni where $D_1$ is the unit disk and the coordinates 
$\c,\c^*$ are related to the
stereographical projection of the hyperboloid on the disk. The inverse
transformation is:
\be
z_1=\sqrt{\f{1}{1-\c\c^*}}\eta,\;\; z_2=
\sqrt{\f{1}{1-\c\c^*}}\c\eta. \label{zs}
\ee

The group law $U''=U'U$ in $\eta,\c,\c^*$ coordinates is:
\begin{eqnarray}
\eta'' &=&\frac{z_1''}{|z_1''|}=\frac{\eta'\eta+\eta'\eta^*\c'\c^*}{\sqrt{(1+
{\eta^*}^2\c'\c^*)(1+\eta^2\c{\c^*}')}}  \nn \\
\c''&=&\frac{z_2''}{z_1''}=\frac{\c\eta^2+\c'}{\eta^2+\c'\c^*}  \\
{\c^*}''&=&\frac{{z_2''}^*}{{z_1''}^{*}}=
\frac{\c^*\eta^{-2}+{\c^*}'}{\eta^{-2}+
{\c^*}'\c},  \nn
\end{eqnarray}

\ni from which we can extract the left- and right-invariant vector fields:
\bea
\XL_\eta &=&\eta\f{\p}{\p\eta}-2\c\f{\p}{\p \c}+2\c^*
\f{\p}{\p \c^*}\label{campos}   \\
\XL_\c &=&-\f{1}{2}\eta \c^* \f{\p}{\p\eta}+
\f{\p}{\p \c}-{\c^*}^2\f{\p}{\p \c^*}\nn \\
\XL_{\c^*} &=&\f{1}{2}\eta \c \f{\p}{\p\eta}-
\c^2 \f{\p}{\p \c}+\f{\p}{\p \c^*}\nn \\
\XR_\eta &=&\eta \f{\p}{\p\eta} \nn \\
\XR_\c &=& \f{1}{2}\eta^{-1} \c^* \f{\p}{\p\eta}+\eta^{-2}(1-\c\c^*)
\f{\p}{\p \c}\nn \\
\XR_{\cc} &=&- \f{1}{2}\eta^3 \c \f{\p}{\p\eta}+\eta^{2}(1-\c\c^*)
\f{\p}{\p\cc}.\nn
\eea
\ni They close the Lie algebra:
\be
\l[\XL_\eta,\XL_\c\r] = 2\XL_\c \,,\,\,\,
\l[\XL_\eta,\XL_{\cc}\r] = -2\XL_{\cc} \,,\,\,\,
\l[\XL_\c,\XL_{\cc}\r] = \XL_\eta\,, \label{conmutadores}
\ee
\ni and the corresponding right version by changing the sign to the structure
constants.

Let us parameterize $G=SO(2,2)$ as two copies of $SU(1,1)$ with
parameters $\{(\eta,\c,\cc);\\ (\heta,\hc,\hcc)\}$. 
There are two possibilities of
combining the generators in the Lie
algebra 
\be
{\cal G}^L=\{\XL_\eta, \XL_\c, \XL_{\cc}, \XL_{\heta}, \XL_{\hc}, 
\XL_{\hcc}\}
\ee
\ni of $G$,
in order to get the usual conformal
generators 
\be 
{\cal G}^L=\{D^L, M^L, {P_0}^L, {P_1}^L, {K_0}^L, {K_1}^L\}
\ee
\ni which 
fulfil the ordinary commutation relations \cite{Kastrup}:
\be
\begin{array}{lll}
\l[\POL,\DL \r]=-\POL & \l[\PL,\DL \r]=-\PL & 
\l[\POL,\ML \r]=-\PL   \\  \l[\PL,\ML \r] =-\POL & \l[\POL,\KOL \r]=-2\DL & 
\l[\PL,\KOL \r]=-2\ML  \\
\l[\POL,\KL \r]=2\ML & \l[\PL,\KL \r]=2\DL & \l[\KOL,\DL \r]=\KOL   \\
 \l[\KL,\DL \r]=\KL & \l[\KOL,\ML \r] =-\KL & \l[\KL,\ML \r] =-\KOL  \\
 \l[\DL,\ML \r]=0 & \l[\POL,\PL \r]=0 & \l[\KOL,\KL \r] =0 
\end{array}\label{conforme}
\ee
\ni where $D,M,P_\mu,K_\mu$ are the generators of dilatation, 
boosts, space-time
translations and special conformal transformations, respectively. One of the
two mentioned choices lead to a non-compact dilatation subgroup, whereas the
other leads to a compact one. Let us show what this combinations are in both
cases:
\be
\begin{array}{ll} \hbox{COMPACT D} & \hbox{NON COMPACT D}  \\
\DL=-\f{1}{2}\left(\XL_\eta+\XL_{\heta}\right) &
\DL=-\f{i}{2}\left(\XL_\c-\XL_{\cc}+\XL_{\hc}-\XL_{\hcc}\right) \\
\ML=\f{1}{2}\left(\XL_\eta-\XL_{\heta}\right) &
\ML=\f{i}{2}\left(\XL_\c-\XL_{\cc}-\XL_{\hc}+\XL_{\hcc}\right) \\
\POL=-\left(\XL_{\cc}+\XL_{\hcc}\right)&
\POL=\f{1}{2}\left(\XL_\c+\XL_{\cc}-\XL_{\hc}-\XL_{\hcc}-
i(\XL_\eta-\XL_{\heta})
\right) \\
\PL=\XL_{\cc}-\XL_{\hcc}&
\PL=-\f{1}{2}\left(\XL_\c+\XL_{\cc}+\XL_{\hc}+\XL_{\hcc}-
i(\XL_\eta+\XL_{\heta})
\right) \\
\KOL=\XL_\c+\XL_{\hc}&
\KOL=\f{1}{2}\left(-\XL_\c-\XL_{\cc}+\XL_{\hc}+\XL_{\hcc}-
i(\XL_\eta-\XL_{\heta})
\right) \\
\KL=\XL_\c-\XL_{\hc}&
\KL=-\f{1}{2}\left(\XL_\c+\XL_{\cc}+\XL_{\hc}+\XL_{\hcc}+
i(\XL_\eta+\XL_{\heta})
\right)\end{array}\label{nonyes}
\ee

The group $G=SU(1,1)\otimes SU(1,1)$ is non-compact and semisimple. The 
left-invariant integration volume can be expressed as: 
\be
v(g)\equiv \theta^L_{g^i}\wedge\stackrel{\hbox{dim}(G)}{...}
\wedge\theta^L_{g^n}=-\f{1}{\eta(1-\c\cc)^2}
\f{1}{\heta(1-\hc\hcc)^2}d\c\wedge d\cc\wedge d\eta\wedge d\hc\wedge d\hcc
\wedge d\heta, \label{volume}
\ee
\ni which becomes singular for values $|\c|,|\hc|\rightarrow 1$ 
(unit circumferences surrounding both open disks). However, resorting to
a central extension $\tilde{G}$ of $G$, necessarily trivial since $G$ is
semisimple and finite-dimensional, we shall turn the extended scalar product, 
between wave functions on the group, finite for some range of the extension 
parameter. 

There are several central extensions of the conformal group, 
but we are interested 
in one that afterwards leads to a generalized Minkowski space. This choice 
corresponds to an extension by a coboundary locally generated by 
the dilatation 
parameter, which we shall consider as a  ``proper time" 
(see Ref. \cite{VAL}).

We shall separate the two cases:
a) non-compact and b) compact dilatation subgroup, in two subsections,
respectively. 
The essentials of the problem 
we are involved in are insensitive to the topological 
character of the dilatation subgroup; however, whereas the non 
compact dilatation case will be useful to connect 
with some standard expressions 
in Minkowski space, the compact dilatation case will be more 
manageable to construct and illustrate the second quantization program.
 It can be proven that a consistent quantum theory needs 
the group $C^*$ as the structure group $T$ for the first case, 
whereas a pseudo-extension 
by $U(1)$ is enough for the second one. 

\subsection{Non-compact dilatation subgroup}

Let us look for a  $T=C^*=\left\{z=\ere\zeta;\,\ere\in\Re^+,\,\zeta\in 
U(1)\right\}$-pseudo-extension
\be
\z''=\z'\z
e^{\xi(g',g)},\;\;\;\;\xi(g',g)=\delta(g'*g)-
\delta(g')-\delta(g),\;\;\;\;\z \in C^*
\ee
\ni where $\delta(g)=-i\beta(\c-\cc+\hc-\hcc)$ is the function which generates
the coboundary and $\beta={\beta_1}+i{\beta_2}$ is a 
complex parameter characterizing 
the representation. 

 The extended left- and right-invariant vector fields in $\tilde G$
are

\bea
\TXL_\ere&=&\TXR_\ere=\ere\f{\p}{\p\ere}\nn \\
\TXL_\zeta&=&\TXR_\zeta=\zeta\f{\p}{\p\zeta}\nn \\
\TXL_\eta&=&\XL_\eta+2i{\beta_1}(\c+\cc)
\TXL_\ere-2{\beta_2}(\c+\cc)\TXL_\zeta\nn\\ 
\TXL_\c&=&\XL_\c-i{\beta_1}{\cc}^2\TXL_\ere+{\beta_2}{\cc}^2\TXL_\zeta\nn\\
\TXL_{\cc}&=&\XL_{\cc}+i{\beta_1} \c^2\TXL_\ere-{\beta_2} \c^2\TXL_\zeta\nn\\
\TXR_\eta&=&\XR_\eta  \nn\\
\TXR_\c&=&\XR_\c-i{\beta_1}(\eta^{-1}(1-\c\cc)-1)\TXR_\ere+
{\beta_2}(\eta^{-1}(1-\c\cc)-1)\TXR_\zeta \nn  \\
\TXR_{\cc}&=&\XR_{\cc}+i{\beta_1}(\eta^{2}(1-\c\cc)-1)\TXR_\ere
-{\beta_2}(\eta^{2}(1-\c\cc)-1)\TXR_\zeta
 \label{ecampos}
\eea
\ni and similar expression for the $\heta,\hc,\hcc$ parameters. The new
commutation relations for the extended conformal Lie algebra $\tilde{\cal G}$
of $\tilde G$ are two copies of:
\bea
\l[\TXL_\eta,\TXL_\c\r]&=& 2\TXL_\c-2i{\beta_1}\TXL_\ere+2{\beta_2}
\TXL_\zeta \nn \\
\l[\TXL_\eta,\TXL_{\cc}\r] &=& -2\TXL_{\cc}-2i{\beta_1}\TXL_\ere+
2{\beta_2}\TXL_\zeta\nn \\
\l[\TXL_\c,\TXL_{\cc}\r] &=& \TXL_\eta \nn \\
\l[\TXL_\ere,\hbox{all}\r]&=&0\nn\\
 \l[\TXL_\zeta,\hbox{all}\r]&=&0\,\,, \label{econmutadores}
\eea
\ni the right ones changing a global sign in the structure constants. 
The only change
induced in the Lie algebra commutators, when expressed in terms of
\be
\tilde{{\cal G}}^L=\{\TDL,\TML,\TPOL,\TPL,\TKOL,\TKL,\TXL_\zeta,\TXL_\ere\}
\ee
\ni (having the
same functional form as
in the right hand side of eq. (\ref{nonyes})), is in the following two
commutators:
\bea
\l[\TPOL,\TKOL\r]&=& -2\TDL+4{\beta_1}\TXL_\ere+4i{\beta_2}\TXL_\zeta \nn \\
\l[\TPL,\TKL\r]&=& 2\TDL-4{\beta_1}\TXL_\ere-4i{\beta_2}\TXL_\zeta\,\,,    
\label{conjugatet}
\eea
\ni the remainder keeping the same expression as in Eq. (\ref{conforme}). These
relations show that the two couples of generators $\TPOL,\TKOL$ and $\TPL,\TKL$
are canonically conjugate, i.e. they give rise to central terms at the 
right-hand side of the corresponding commutator. Central extensions of 
this kind were 
already considered in Refs. \cite{Bisquert,VAL}.   
 From (\ref{conjugatet}) we conclude 
that, like the space operator, time is not deprived of 
dynamical character, that is, 
it is an operator subject to uncertainty relations (see \cite{franchutes} 
for another definition of space-time position operators inside the enveloping 
algebra of the conformal group).

The quantization form and its characteristic module are
\bea
\Theta&=&\left(\Theta^{(\ere)},\Theta^{(\zeta)}\right)\nn\\
\Theta^{(\ere)}&=&{\beta_1}(\Gamma(\eta,\c,\cc)+\Gamma(\heta,\hc,\hcc))+
\ere^{-1}d\ere\nn\\
\Theta^{(\zeta)}&=&-{\beta_2}(\Gamma(\eta,\c,\cc)+\Gamma(\heta,\hc,\hcc))+
i\zeta^{-1}d\zeta \nn\\
\Gamma(\eta,\c,\cc)&\equiv&\f{1}{1-\c\cc}\left(-2i(\c+\cc)\eta^{-1}d\eta
-i\c\cc d\c+i\c\cc d\cc\right)\nn\\
{\cal G}_\Theta &=& <\TDL,\TML>.
\eea

Let ${\cal B}(\TG)$ be the set of complex valued
$T$-{\it functions} on $\TG$ in the sense
of principal bundle theory: $\psi(\z *\tg)=D_T(z)\psi(\tg)$ and let 
us choose the representation $D_T(z)=z^p$, where $p$ has to be a 
negative integer for single-valuedness  and ``square integrable" 
condition of the wave function. In order to reduce the representation of 
$\TG$ on  ${\cal B}(\TG)$, we impose the  full
polarization subalgebra:
\be
{\cal P}=<\TDL,\TML,\TKOL,\TKL>\,.\label{polsub}
\ee
\ni  The solution to the polarization conditions leads to a Hilbert space 
${\cal H}(\TG)$ made of wave functions of the 
form

\bea
\psi^{(\beta)}(\eta,\c,\cc,\heta,\hc,\hcc,\z)&=&\z^p W_\beta(\c,\cc,\hc,\hcc)
\phi(\mu,\hmu) \nn \\
W_\beta(\c,\cc,\hc,\hcc)&=&w_\beta(\c,\cc)w_\beta(\hc,\hcc)\nn \\
w_\beta(\c,\cc)&=&(1-\c\cc)^{p\beta}(\c+i)^{-p\beta}(\cc-i)^{-p\beta}
e^{ip\beta(\c-\cc)}\,, \label{pol}
\eea
\ni where $W_\beta$ is a ``generating function" and $\phi$ is an arbitrary 
power series
\be
\phi(\mu,\hmu)=\sum_{n,\hn=-\infty}^{\infty}
a_{n,\hn}\phi_{n,\hn}(\mu,\hmu),\;\;\;\phi_{n,\hn}(\mu,\hmu)\equiv 
\mu^n{\hmu}^\hn
\ee
\ni in the variables
\be
\mu=\f{\cc-i}{\c+i}\eta^{-2}=\f{{z_2^*}-i{z_1^*}}{{z_2}+i{z_1}},\;\;\;\;
\hmu=\f{\hcc-i}{\hc+i}\heta^{-2}=\f{{\hzc}_2-i{\hzc}_1}{{\hz}_2
+i{\hz}_1}\,.
\ee
\ni Note that $(\mu,\hmu)$ are defined in a two-dimensional torus 
$T^2=S^1\times S^1$ (the 1+1 dimensional version of the 3+1 
dimensional projective cone $S^3\times S^1/Z_2$). Let
us show how the conformal group act on $T^2$ {\it free from singularities}. 
For this, we  have only to translate the group composition law, originally 
written in global variables $z_i, \hz_i,\, i=1,2$, in Eq.(\ref{zs}), 
to the variables $\mu,\hmu$:
\bea
\mu &\rightarrow&  \mu''\equiv\f{{z_2^*}''-i{z_1^*}''}{{z_2}''+i{z_1}''}=
\f{{z_2^*}'{z_2}+{z_1^*}'{z_1^*}-i({z_2^*}'{z_1}+{z_1^*}'{z_2^*})}{{z_1}'
{z_2}+{z_2}'{z_1^*}+i({z_1}'{z_1}+{z_2}'{z_2^*})}=
\f{\mu-i\cc{}'}{{\eta'}^2(1+i\mu \c')} \nn \\
\hmu &\rightarrow&  \hmu{}''\equiv\f{{\hzc}_2{}''-i{\hzc}_1{}''}{{\hz}_2{}''
+i{\hz}_1{}''}=
\f{{\hzc}_2{}'{\hz}_2+{\hzc}_1{}'{\hzc}_1{}-i({\hzc}_2{}'
{\hz}_1{}+{\hzc}_1{}'{\hzc}_2{})}{{\hz}_1{}'{\hz}_2{}+{\hz}_2{}'{\hzc}_1{}
+i({\hz}_1{}'{\hz}_1{}+{\hz}_2{}'{\hzc}_2{})}=
\f{\hmu-i\hcc{}'}{\heta'{}^2(1+i\hmu {\hc}')}\label{singfree}\,\,.
\eea
\ni This action is always well defined and transitive on $T^2$ (see Ref. 
\cite{Mack} for a more detailed study of the global properties of a similar 
space in 3+1 dimensions), in contrast to 
the action on the Minkowski space, which can
be seen as a local chart of $T^2$
obtained by stereographical projection ($\mu\equiv 
e^{i\theta},\,\hmu\equiv e^{i\bar{\theta}}$):
\bea
t&=&\f{1}{2}(\cot\f{\theta}{2}+\cot\f{\bar{\theta}}{2})\nn\\
x&=&\f{1}{2}(\cot\f{\theta}{2}-\cot\f{\bar{\theta}}{2}),
\eea
\ni as can be checked by realizing that the expression of the 
 generators of  the conformal group in $T^2$ (see Eq. 
(\ref{conftoro})) acquire the standard form in Minkowski space 
--except for (quantum) inhomogeneous terms proportional to the extension 
parameter $\beta$-- (see \cite{Kastrup} for instance) 
when expressed in terms of $t,x$. The manifold 
$T^2$ is thus a natural space-time on which a globally-defined 1+1 conformaly 
invariant QFT can live.

 The invariant integration volume is $v(\tg)=v(g)\wedge
(\ere^{-1}d\ere)\wedge(i\zeta^{-1}d\zeta)$ (see Eq.(\ref{volume})). 
The scalar product of two wave functions (\ref{pol}) will be finite
when the factor $((1-\c\cc)(1-\hc\hcc))^{2p\beta}$,  coming
from  $W_\beta$ (see Eq.(\ref{pol})), cancels out the singularity of
$v(\tg)$ at the boundary of the unit disk due to the factor
$((1-\c\cc)(1-\hc\hcc))^{-2}$. This is  possible when
\be
p{\beta_1} > 1/2 \label{acotada}\;,
\ee
\ni with no restriction in the parameter ${\beta_2}$ (this is the reason why 
the pseudo-extension by the real positive line, with parameter 
${\beta_1}\not=0$, is fundamental for this case).

The action of the right-invariant vector fields (operators in the theory) 
on polarized wave functions (see Eq. (\ref{pol})) has the explicit form:
\bea
\TDR\psi^{(\beta)} &=&\z^p W_\beta\cdot 
\left(-\f{1}{2}(\mu^2-1)\frac{\p}{\p\mu}-\f{1}{2}(\hmu^2-1)
\frac{\p}{\p\hmu}-p\beta(\mu+\mu^{-1}+\hmu+
\hmu^{-1}-2)\right)\phi(\mu,\hmu)\nn\\
\TMR\psi^{(\beta)}&=&\z^p W_\beta\cdot
\left(\f{1}{2}(\mu^2-1)\frac{\p}{\p\mu}-\f{1}{2}(\hmu^2-1)\frac{\p}{\p\hmu}
-p\beta(-\mu-\mu^{-1}+\hmu+\hmu^{-1})\right)\phi(\mu,\hmu)\nn\\
\TPOR\psi^{(\beta)}&=&\z^p W_\beta\cdot
\left(-\f{i}{2}(\mu-1)^2\frac{\p}{\p\mu}+\f{i}{2}(\hmu-1)^2\frac{\p}{\p\hmu}
-p\beta(\mu-\mu^{-1}-\hmu+\hmu^{-1})\right)\phi(\mu,\hmu)\nn\\
\TPR\psi^{(\beta)}&=&\z^p W_\beta\cdot\left(\f{i}{2}(\mu-1)^2
\frac{\p}{\p\mu}+\f{i}{2}(\hmu-1)^2\frac{\p}{\p\hmu}
-p\beta(-\mu+\mu^{-1}-\hmu+\hmu^{-1})\right)\phi(\mu,\hmu)\nn\\
\TKOR\psi^{(\beta)}&=&\z^p W_\beta\cdot\left(\f{i}{2}(\mu+1)^2
\frac{\p}{\p\mu}-\f{i}{2}(\hmu+1)^2\frac{\p}{\p\hmu}
+p\beta(\mu-\mu^{-1}-\hmu+\hmu^{-1})\right)\phi(\mu,\hmu)\nn\\
\TKR\psi^{(\beta)}&=&\z^p W_\beta\cdot\left(\f{i}{2}(\mu-1)^2
\frac{\p}{\p\mu}+\f{i}{2}(\hmu-1)^2\frac{\p}{\p\hmu}
-p\beta(-\mu+\mu^{-1}-\hmu+\hmu^{-1})\right)\phi(\mu,\hmu)\nn\\
\TXR_\ere\psi^{(\beta)}&=&p\psi^{(\beta)},\,\,\TXR_\zeta\psi^{(\beta)}=
p\psi^{(\beta)}\label{conftoro}.
\eea

\ni This representation is irreducible for the extended
conformal group $\TG$ and this is a consequence, according to the general
formalism, of the maximality of the full polarization subalgebra ${\cal P}$
in Eq. (\ref{polsub}), i.e. ${\cal P}$ cannot be further enlarged nor the 
representation further reduced. The process of obtaining the 
unitary irreducible 
representations ends here. Any restriction desired on our wave functions 
should then be imposed as constraints.

We are interested, however, in null mass representations, and these 
can be achieved by selecting those wave functions $\psi^{(\beta)}_c$ in 
${\cal H}(\TG)$ 
which are nullified  by the Casimir $\TQR\equiv({\TPOR})^2-({\TPR})^2$ of the 
Poincar\'e subgroup. More explicitly, wave functions which fulfil:
\bea
\TQR\psi^{(\beta)}_c=0&\Rightarrow&\left(\f{(\hmu-1)^2}{(\hmu-\hmu^{-1})}
\f{\p}{\p\hmu}+p\beta\right)\left(\f{(\mu-1)^2}{(\mu-\mu^{-1})}
\f{\p}{\p\mu}+p\beta\right)\phi(\mu,\hmu)=0\nn\\
&\Rightarrow&\f{\p\varphi(\mu,\hmu)}{\p\mu\p\hmu}=0\,, \label{ligadurakg}
\eea
\ni where
\be
\phi(\mu,\hmu)\equiv 
\left(\f{(\mu-1)^2}{\mu}\f{(\hmu-1)^2}{\hmu}\right)^{-p\beta}
\varphi(\mu,\hmu).
\ee
\ni This Klein-Gordon-like evolution equation 
(in a light-cone-like coordinates) is then 
interpreted as a constraint in the theory and leads to a new Hilbert space 
${\cal H}_c(\TG)$ made of constrained wave functions of the form:
\be
\psi^{(\beta)}_c=z^pW_\beta\left(\f{(\mu-1)^2}{\mu}
\f{(\hmu-1)^2}{\hmu}\right)^{-p\beta}
(\varphi(\mu)+\bar{\varphi}(\hmu)),\label{kgm}
\ee
\ni that is, wave functions for which the arbitrary part splits up into 
functions which depend on $\mu$ and 
$\hmu$ separately (they resemble the standard left- and right-hand 
moving modes).
So long as this constraint is imposed by means of generators of the left 
translation on the group, not all the 
operators ${\TXR}_{g^i}$ will preserve this constraint; only the ones called
{\it good} in the general approach of Refs. \cite{Ramirez,FracHall} will do. 
One can obtain the 
good operators for the condition (\ref{ligadurakg}) by looking at the 
(right) commutators:
\bea
\l[\TDR,\TQR\r]&=&-2\TQR \nn \\
\l[\TMR,\TQR\r]&=&0 \nn \\
\l[\TPOR,\TQR\r]&=&0\nn \\
\l[\TPR,\TQR\r]&=&0\nn \\
\l[\TKOR,\TQR\r]&=&-4\TPOR\TDR+4\TPR\TMR-8ip\beta\TPOR \nn \\
&=& f_0(\mu,\hmu)\TQR-8ip\beta\TPOR\nn\\
\l[\TKR,\TQR\r]&=&-4\TPR\TDR+4\TPOR\TMR-8ip\beta\TPR\nn\\
&=& f_1(\mu,\hmu)\TQR-8ip\beta\TPR\,,\label{goodcom}
\eea
\ni [$f_\nu(\mu,\hmu)$ are some functions on the torus], from which 
we can conclude  that the set of (first-order) good operators is

\be
{\cal G}_{good}=<\TDR,\TMR,\TPOR,\TPR,\TXR_\ere,\TXR_\zeta>,
\ee
\ni and close a subalgebra (Poincare$+$dilatation$\equiv$Weyl)
of the extended conformal Lie algebra in $1+1$ dimensions.

The fact that  $\TKOR$ and $\TKR$ are {\it bad} operators, i.e. they do not 
preserve the ${\cal H}_c(\TG)$ Hilbert space, will be relevant in the 
``second quantization" of the constrained theory.  The new (Weyl) vacuum will 
no longer be annihilated by the second quantized version of  
$\TKOR$ and $\TKR$  but, rather, it will appear 
to be ``polarized'' from an accelerated frame (see Subsec. 4.2).
This way, the profound reason for the
rearrangement of the vacuum (under special conformal transformations) in
(massless) Quantum Conformal Field Theories is not a singular 
action of this subgroup
on the space-time but, rather, the impossibility of properly implementing 
these transformations in the constrained Hilbert space ${\cal H}_c(\TG)$. 
Note that the combinations $A_+\equiv\frac{1}{2}(\TKOR+\TKR)$ and  
$A_-\equiv\frac{1}{2}(\TKOR-\TKR)$ 
are ``partially good'', in the sense that they preserve 
the left- and right-hand moving modes subspaces, respectively; we shall see 
(Subsec. 4.2) how  its finite 
action on a Weyl vacuum (in the second quantized theory)  
give rise to a thermal bath of left- and right-hand moving scalar photons, 
respectively.

As far as the classical field theories is concerned, the existence of a well 
defined scalar product does not really matter; the condition (\ref{acotada}) 
can be violated by putting $\beta=0$, thus leading to a reducible 
representation where the operators $\TKOR$ and
$\TKR$ leave  the equation $\TQR\psi^{(\beta)}_c=0$ invariant, 
as it can be easily checked 
from the two last commutators in (\ref{goodcom}). However, for this 
particular case, the loss of unitarity can give rise to some problems in 
the quantization procedure, especially concerning 
the definition of the field propagators in the quantum field theory 
(see Sec. 4). Thus, for the null mass case, the conformal symmetry is 
``spontaneously broken" in the sense that it is a symmetry of the classical 
massless field theory, whereas the corresponding quantum field theory is only 
invariant under the Weyl subgroup. The appearance of terms proportional to 
$\beta$ at the right hand side of some commutators, as in (\ref{goodcom}), 
can be seen as an ``anomaly" ; however, this time, anomaly does not means 
obstruction to quantization but, on the contrary, it is intrinsic to the 
quantization procedure and necessary for the good behaviour of the theory.

Note that for {\it massive} field theories the situation is 
slightly different. 
The only symmetry which survives (both for classical and quantum theories), 
after  the constraint 
\be
\TQR\psi^{(\beta)}_c=D^{(m)}(\TQR)\psi^{(\beta)}_c=
m^2\psi^{(\beta)}_c \label{massive}
\ee
is imposed, is the Poincar\'e subgroup. Indeed, the dilatation generator 
is now a bad operator (it does not preserve the constraint (\ref{massive}), as 
can be seen from the first line of (\ref{goodcom})). 
Its finite action, of course 
being bad, is not ``so bad" in the sense that it changes from one 
representation $D^{(m)}(\TQR)$ to another $D^{(m')}(\TQR)$ with 
$m'=e^{2\lambda}m$, where $\lambda$ is the parameter of the transformation. 
That is, it plays the role of a ``quantization-changing operator" (see 
Ref. \cite{FracHall} for other relevant examples), its domain being the 
union $\bigcup_{m\in\Re^+}{\cal H}^{(m)}_c(\TG)$ of all the constrained 
Hilbert spaces corresponding to different masses (i.e. a theory with continuum 
mass spectrum). 

 One can look for a physical interpretation 
of those facts and say that ``quantum
conformal fields  do not evolve in time". The representation
(\ref{conftoro}) is irreducible for the whole conformal group, but reducible
under Poincare$+$dilatation (Weyl) subgroup.
Some external perturbation breaks the conformal symmetry and forces the fields
to evolve in time and acquire a fixed value for the mass 
(we are interested in the massless case), so that these fields carry an 
irreducible  representation of the Poincare($+$dilatation) subgroup. 
In this way, the dynamical
symmetry breaking and the fixing of the mass, even null, come together.
\vskip 0.5cm

\subsection{Compact dilatation subgroup}

\vskip 0.5cm
It can be proved that, for this case,  a $T=U(1)$-pseudo-extension 
is enough to have a well defined quantum theory. It has the form:
\be
\zeta''=\zeta'\zeta
e^{i\xi(g',g)}=\zeta'\zeta\left(\eta''{\eta'}^{-1}\eta^{-1}\heta''
{\heta'}{}^{-1}\heta^{-1}\right)^{-2N}\,,\label{cdp}        
\ee
\ni where $\xi (g',g)$ is the two-cocycle (in fact, coboundary) generated
by a multiple of $i(\log\eta+\log\heta)$, and the parameter $N$ 
labels the irreducible  
representations and it must  be quantized, taking the values
\be
N=\f{j}{2},\;\;\;\; j\in Z,\label{acotada2}
\ee
\ni for globality conditions.  

 The extended left- and right-invariant vector fields on $\tilde G$
are:

\be
\begin{array}{ll}\TXL_\eta=\XL_\eta &\TXR_\eta=\XR_\eta  \\
\TXL_\c=\XL_\c+N\cc\TXR_\zeta&
\TXR_\c=\XR_\c-N\eta^{-2}\cc\TXR_\zeta  \\
\TXL_{\cc}=\XL_{\cc}-N\c\TXL_\zeta &\TXR_{\cc}=\XR_{\cc}+N
\eta^{2}\c\TXR_\zeta\end{array} \label{cecampos}
\ee
\ni and similar expressions for the $\heta,\hc,\hcc$ parameters. The new
commutation relations for the extended conformal Lie algebra $\tilde{\cal G}$
of $\tilde G$ are two copies of:
\bea
\l[\TXL_\eta,\TXL_\c\r]&=& 2\TXL_\c \nn \\
\l[\TXL_\eta,\TXL_{\cc}\r] &=& -2\TXL_{\cc} \nn \\
\l[\TXL_\c,\TXL_{\cc}\r] &=& \TXL_\eta -2N\TXL_\zeta\nn \\
\l[\TXL_\zeta,\hbox{all}\r]&=&0. \label{ceconmutadores}
\eea
\ni  which, expressed in terms of the basis
$\{\TDL,\TML,\TPOL,\TPL,\TKOL,\TKL,\TXL_\zeta\}$,
 lead now to
\bea
\l[\TPOL,\TKOL\r]&=& -2\TDL-4N\TXL_\zeta \nn \\
\l[\TPL,\TKL\r]&=& 2\TDL+4N\TXL_\zeta    \label{confcomp}
\eea
\ni and the same expression as in (\ref{conforme}) for the 
remainder.

 The left-invariant 1-form $\Theta$ 
has  now the form:
\bea
\Theta&=&\f{iN}{1-\c\cc}\left(4\c\cc\eta^{-1}d\eta+\cc dc-\c
d\cc\right)\nn\\
&+&\f{iN}{1-\hc\hcc}\left(4\hc\hcc\heta^{-1}d\heta+\hcc d\hc-\hc
d\hcc\right)-i\zeta^{-1}d\zeta,  
\eea
\ni the characteristic module ${\cal G}_\Theta$ and the 
polarization subalgebra 
having the same content in fields as in the previous section.  
The polarized $U(1)$-functions (we choose the faithful 
representation for $U(1)$) have now the form
\bea
\psi^{(N)}(\eta,\c,\cc,\heta,\hc,\hcc,\zeta)&=&\zeta W_N(\c,\cc,\hc,\hcc)
\phi(s,\hs) \nn \\
W_N&=&w_N(\c,\cc)w_N(\hc,\hcc)\nn \\
w_N(\c,\cc)&=&(1-\c\cc)^{N}       \label{cpol}
\eea
\ni where $W_N$ is a ``generating function'' 
and $\phi$ is an arbitrary power series
\be
\phi(s,\hs)=\sum_{n,\hn=0}^{\infty}
a_{n,\hn} s^n{\hs}^\hn
\ee
\ni in the variables
\be
s=\eta^{-2}\cc=\f{{z_2^*}}{{z_1}},\;\;\;\;\hs=\heta^{-2}\hcc=
\f{{\hzc}_2{}}{{\hz}_1{}}     .
\ee

Let us show how the conformal group act on $s,\hs$ free from singularities. 
For this, let us  proceed as in Eq. (\ref{singfree}):
\bea
s &\rightarrow&  s''\equiv\f{{z_2^*}''}{{z_1}''}=
\f{{z_2^*}'{z_2}+{z_1^*}'{z_1^*}}{{z_1}'{z_1}+{z_2}'{z_2^*}}=
\f{s+\cc{}'}{{\eta'}^2(1+s \c')} \nn \\
\hs &\rightarrow&  \hs{}''\equiv\f{{\hzc}_2{}''}{{\hz}_1{}''}=
\f{{\hzc}_2{}'{\hz}_2{}+{\hzc}_1{}'{\hzc}_1{}}{{\hz}_1{}{\hz}_1{}+{\hz}_2{}'
{\hzc}_2{}}=
\f{\hs+\hcc{}'}{\heta'{}^2(1+\hs \hc')}
\eea
\ni This action is always well defined and transitive on this space.

The invariant integration volume can be now chosen as $v(\tg)=-
(2\pi)^{-5}v(g)\wedge(i\zeta^{-1}d\zeta)$ and the scalar product 
of two basic functions ${\check{\psi}}^{(N)}_{n,\hn}\equiv\zeta 
W_N s^n{\hs}^\hn$ and ${\check{\psi}}^{(N)}_{m,\hm}\equiv\zeta 
W_N s^m{\hs}^\hm$ is:
\bea
\langle \check{\psi}^{(N)}_{n,\hn}|\check{\psi}^{(N)}_{m,\hm}\rangle &=& 
\f{n!(2N-2)!}{(2N+n-1)!}\f{\hn!(2N-2)!}{(2N+\hn-1)!}
\delta_{nm}\delta_{\hn\hm}
= C^{(N)}_nC^{(N)}_\hn\delta_{nm}\delta_{\hn\hm}\nn\\
 C^{(N)}_n &\equiv&\f{n!(2N-2)!}{(2N+n-1)!},\label{acotada3}
\eea
\ni where we are assuming that $N > \f{1}{2}$, a necessary 
condition for having a well defined (finite) scalar product 
[this condition can be relaxed to $N> 0$ by going to the 
universal covering group of $G$]. 
The set 
\be 
B({\cal H}_N(\TG))=\left\{ \psi^{(N)}_{n,\hn}\equiv
\f{1}{\sqrt{C^{(N)}_nC^{(N)}_\hn}}
\check{\psi}^{(N)}_{n,\hn} \right\}\label{basicwave}
\ee  
is then orthonormal and complete, i.e. an orthonormal base of 
${\cal H}_N(\TG)$.

The actions of the right-invariant vector fields (operators in the theory) on
polarized wave functions (see Eq. (\ref{cpol})) have the explicit form:
\bea
\TDR\psi^{(N)}&=&\zeta W_N\cdot(s\f{\p}{\p s}+\hs\f{\p}{\p\hs})\phi(s,\hs)
\nn\\
\TMR\psi^{(N)}&=&\zeta W_N\cdot(-s\f{\p}{\p s}+\hs\f{\p}{\p\hs})\phi(s,\hs)
\nn\\
\TPOR\psi^{(N)}&=&\zeta W_N\cdot(-\f{\p}{\p s}-\f{\p}{\p\hs})\phi(s,\hs)
\nn\\
\TPR\psi^{(N)}&=&\zeta W_N\cdot(\f{\p}{\p s}-\f{\p}{\p\hs})\phi(s,\hs)
\nn\\
\TKOR\psi^{(N)}&=&\zeta W_N\cdot(-s^2\f{\p}{\p s}-\hs^2\f{\p}{\p\hs}-
2N(s+\hs))\phi(s,\hs)\nn\\
\TKR\psi^{(N)}&=&\zeta W_N\cdot(-s^2\f{\p}{\p s}+\hs^2\f{\p}{\p\hs}-
2N(s-\hs))\phi(s,\hs)\nn\\
\TXL_\zeta\psi^{(N)}&=&\psi^{(N)}.
\eea

The finite (left) action (\ref{leftaction}) of an arbitrary element 
$\tg'=(\eta',\c',{\cc}',\heta',\hc',{\hcc}{}',\zeta')\in\TG$ on an 
arbitrary wave function 
\be
\psi^{(N)}(\tg)=\sum^\infty_{n,\hn=0}a_{n,\hn}\psi^{(N)}_{n,\hn}(\tg),
\ee
\ni  can 
be given through the matrix elements $\A(\tg')\equiv \langle 
\psi^{(N)}_{m,\hm}|\rho(\tg')|\psi^{(N)}_{n,\hn}\rangle$ of $\rho$ in the 
base $B({\cal H}_N(\TG))$. They have the  
following expression:
\bea
\A(\tg)&=&\zeta^{-1}\rho^{(N)}_{mn}(\eta,\c,\cc)
\rho^{(N)}_{\hm\hn}(\heta,\hc,\hcc)\nn\\
\rho^{(N)}_{mn}(\eta,\c,\cc)&=& \sqrt{\f{C^{(N)}_m}{C^{(N)}_n}}
\sum^n_{l=\theta_{nm}}
\left(\begin{array}{c}n\\ l\end{array}\right) 
\left(\begin{array}{c}2N+m+l-1\\ m-n+l\end{array}\right)\times \nn \\
  & &(-1)^l\eta^{2m}{\cc}^l\c^{m-n+l}(1-\c\cc)^N\label{fouritrans}
\eea
\ni where the function $\theta_{nm}$ in the lower limit 
of the last summatory is defined by 
$\theta_{nm}\equiv(n-m)\f{\hbox{sign}(n-m)+1}{2}$, 
the function $\hbox{sign}(n)$ being the standard 
sign function ($\hbox{sign}(0)=1$); it guarantees the following 
inequality $m-n+l\geq 0$. These expressions will be very useful 
for the construction of the corresponding quantum field theory in 
the next section.

The constrained wave functions $\psi^{(N)}_c$ obeying

\be
\TQR\psi^{(N)}_c=(({\TPOR})^2-({\TPR})^2)\psi^{(N)}_c=0\Rightarrow  
\f{\p^2\phi}{\p s\p\hs}=0\label{kglicke}
\ee
\ni have now the form
\be
\psi^{(N)}_c=\zeta W_N \cdot(\varphi(s)+{\bar{\varphi}}(\hs)).
\ee
\ni We arrive at the same conclusions as in the non-compact 
dilatation case, concerning good and bad operators. For this case, 
$N$ plays the same role as $\beta$ did in the former.

Let us investigate the conformal quantum field theory associated 
with this ``first quantized" theory and how to interpret the 
dynamical symmetry breaking of the conformal group in the 
context of the corresponding ``second quantized" theory.
To this end, let us show how this second quantization 
approach can be discussed within  
the  GAQ framework.

\section{``Second Quantization" on a group $\TG$: a model for a conformally 
invariant QFT}

In this subsection we shall develop a general approach to the
quantization of  linear, complex quantum fields defined on a group 
manifold $\TG$ (more precisely, on the quotient $\TG/(T\cup{\cal P})$). 
This formalism can be seen as a ``second quantization" of a ``first quantized" 
theory defined by a group $\TG$ and a Hilbert space ${\cal H}(\TG)$ of 
polarized wave functions. 

The construction of the quantizing group 
$\tilde{G}^{(2)}$ for this complex quantum field is as follows. Given 
a  Hilbert space ${\cal H}(\TG)$ and its dual ${\cal H}^*(\TG)$, we 
define the direct sum 
\bea
{\cal F}(\TG)&\equiv& {\cal H}(\TG)\oplus {\cal H}^*(\TG)\nn \\
&=&\left\{|f\rangle =|A\rangle +|B^*\rangle ;\,\,
|A\rangle \in {\cal H}(\TG),\,|B^*\rangle \in {\cal H}^*(\TG)\right\}\,,
\eea
\ni where we have denoted $|B^*\rangle$ according to 
$\langle \tg^*_\cp|B^*\rangle \equiv 
\langle B|\tg_\cp\rangle =B_\cp^*(\tg)$. The group $\TG$ acts on this 
vectorial space as follows:
\be
\rho(\tg')|f\rangle =\rho(\tg')|A\rangle +\rho(\tg')|
B^*\rangle \,,
\ee
\ni where 
\be
\langle \tg^*_\cp|\rho(\tg')|B^*\rangle \equiv 
\langle B|\rho^{\dag}(\tg')|\tg_\cp\rangle =B_\cp^*(\tg'{}^{-1}*\tg). 
\ee

We can also define the dual space 
\bea
{\cal F}^*(\TG)&\equiv& {\cal H}^*(\TG)\oplus {\cal H}^{**}(\TG)\nn \\
&=&\left\{\langle f|=\langle A|+\langle B^*|\,;\;\;
\langle A|\in {\cal H}^*(\TG),\,\langle B^*|\in {\cal H}^{**}(\TG)\sim 
{\cal H}(\TG)\right\}\,,
\eea
\ni where $\TG$ acts according to the adjoint action
\be
\langle f|\rho^{\dag}(\tg')=\langle A|\rho^{\dag}(\tg')+
\langle B^*|\rho^{\dag}(\tg')
\ee
\ni and now 
\be
\langle B^*|\rho^{\dag}(\tg')|\tg^*_\cp\rangle \equiv
\langle \tg_\cp|\rho(\tg')|B\rangle .
\ee

Using the closure relation (\ref{closure}), the product of two arbitrary 
elements of ${\cal F}(\TG)$ is
\be
\langle f'|f\rangle =\langle A'|A\rangle +
\overbrace{\langle A'|B^*\rangle }^{0}+
\overbrace{\langle B'{}^*|A\rangle }^{0}+
\langle B'{}^*|B^*\rangle \,,
\ee
\ni indeed, the second and third integrals 
\be
\int_{\TG}{ v(\tg)A_\cp'{}^*(\tg)B_\cp^*(\tg)}=0=
\int_{\TG}{ v(\tg)B_\cp'(\tg)A_\cp(\tg)}
\ee
\ni are zero because of the integration on 
the central parameter $\zeta\in U(1)$. Thus, the subspaces 
${\cal H}(\TG)$ and ${\cal H}^*(\TG)$ are orthogonal with respect to this 
scalar product in ${\cal F}(\TG)$. A basis for ${\cal F}(\TG)$ is provided by 
the set  $\left\{|n\rangle + |m^*\rangle \right\}_{n,m\in I}\,$.

The space ${\cal M}(\TG)\equiv{\cal F}(\TG)\otimes{\cal F}^*(\TG)$ 
can be endowed with a simplectic structure
\be
S(f',f)\equiv\f{-i}{2}(\langle f'|f\rangle -
\langle f|f'\rangle )\,,
\ee
\ni thus defining  ${\cal M}(\TG)$ as a phase space. 
This phase space can be naturally embedded into a quantizing group 
\be
\tilde{G}^{(2)}\equiv\left\{\tg^{(2)}=(g^{(2)};\vs)\equiv\left(\tg,|f\rangle ,
\langle f|;\vs\right)\right\}\,,
\ee
\ni which is a (true) central extension by $U(1)$, with parameter $\vs$, of 
the semidirect product $G^{(2)}\equiv\TG\otimes_{\rho}{\cal M}(\TG)$ of 
the basic group $\TG$ and the phase space ${\cal M}(\TG)$. The group 
law of $\tilde{G}^{(2)}$ is formally:
\bea
\tg''&=&\tg'*\tg \nn \\
|f''\rangle &=&|f'\rangle +\rho(\tg')|f\rangle \nn\\
\langle f''|&=&\langle f'|+\langle f|\rho^{\dag}(\tg')\nn\\
\vs''&=&\vs'\vs e^{i\xi^{(2)}(g^{(2)}{}',g^{(2)})}\label{group2}\,,
\eea
\ni where $\xi^{(2)}(g^{(2)}{}',g^{(2)})$ is a two-cocycle defined as
\be
\xi^{(2)}(g^{(2)}{}',g^{(2)})\equiv \kappa S(f',\rho(\tg')f)\,\label{twococy}
\ee
\ni and $\kappa$ is intended to kill any possible dimension of $S$.

A system of coordinates for $\tilde{G}^{(2)}$ corresponds to a choice of 
representation associated with a given polarization ${\cal P}$
\be
\begin{array}{ll} f_\cp^{(+)}(\tg)\equiv\langle \tg_\cp|f\rangle \,, & 
f_\cp^{(-)}(\tg)\equiv\langle \tg_\cp^*|f\rangle \,,\\
f_\cp^{*(+)}(\tg)\equiv\langle f|\tg_\cp^*\rangle \,, &
f_\cp^{*(-)}(\tg)\equiv\langle f|\tg_\cp\rangle \,.\end{array}
\label{confpar}
\ee
\ni This splitting of $f$ is the group generalization of the more standard 
decomposition of a field in positive and negative frequency parts. 
If we make use of the closure relation 
$1 = \int_{\TG} v(\tg)\{ |\tg_\cp\rangle\langle\tg_\cp|  +  
|\tg_\cp^*\rangle\langle\tg_\cp^*| \}$ for ${\cal F}(\TG)$, the explicit 
expression of the cocycle (\ref{twococy}) in this coordinate system 
(for simplicity, we discard the semidirect action of $\TG$),  
\bea 
\xi^{(2)}(g^{(2)}{}',g^{(2)})&=&\f{-i\kappa}{2}\int\!\!\int_{\TG}{
v(\tg')v(\tg) \left\{ 
f_\cp'^{*(-)}(\tg')\Delta_\cp^{(+)}(\tg',\tg)f_\cp^{(+)}(\tg)\right.} \nn\\ 
&-&  f_\cp^{*(-)}(\tg')\Delta_\cp^{(+)}(\tg',\tg)f_\cp'^{(+)}(\tg) 
  + f_\cp'^{*(+)}(\tg')\Delta_\cp^{(-)}(\tg',\tg)f_\cp^{(-)}(\tg)
\label{twococy2} \\ 
&-&  \left. f_\cp^{*(+)}(\tg')\Delta_\cp^{(-)}(\tg',\tg)f_\cp'^{(-)}(\tg) 
\right\}\,,\nn
\eea
\ni where 
\bea
\Delta_\cp^{(+)}(\tg',\tg)&\equiv&\langle\tg_\cp'|\tg_\cp\rangle=
\sum_{n\in I}{
\psi_{\cp,n}(\tg')\psi_{\cp,n}^*(\tg)}\,,\nn\\
\Delta_\cp^{(-)}(\tg',\tg)&\equiv&\langle\tg_\cp'^*|\tg_\cp^*\rangle=
\Delta_\cp^{(+)}(\tg,\tg')\,,\label{propm}
\eea
\ni shows that the vector fields associated with  the co-ordinates in 
(\ref{confpar}) are canonically conjugated
\bea
\l[\TXL_{f^{*(-)}_\cp(\tg')},\TXL_{f_\cp^{(+)}(\tg)}\r]&=&\kappa
\Delta_\cp^{(+)}(\tg',\tg)\TXL_\vs\,,\nn\\
\l[\TXL_{f^{*(+)}_\cp(\tg')},\TXL_{f_\cp^{(-)}(\tg)}\r]&=&\kappa
\Delta_\cp^{(-)}(\tg',\tg)\TXL_\vs\,.
\eea
\ni Here, the functions $\Delta_\cp^{(\pm)}(\tg',\tg)$ play the role of  
{\it propagators} 
(central matrices of the cocycle). At this point, we must stress the 
importance of a well defined scalar product 
in ${\cal H}(\TG)$ as regards the good behaviour of the two-cocycle 
(\ref{twococy2}), an essential ingredient in the corresponding QFT. 
The non-zero 
value of the central extension parameter of $\TG$
--see Eq. (\ref{acotada},\ref{acotada2}) and 
comments after Eq. (\ref{acotada3})--  which prevents the whole conformal 
group from being an exact symmetry of the massless {\it quantum} field theory 
 (remember the comments after Eq. (\ref{goodcom})) proves to be an essential 
prerequisite for a proper definition of the conformal {\it quantum} field 
theory through the group $\TG^{(2)}$.

The propagators in two different 
parametrizations of $\tilde{G}^{(2)}$, corresponding to two different 
polarization subalgebras ${\cal P}_1$ and ${\cal P}_2$ of 
${\tilde{\cal G}}^L$ (or $U{\tilde{\cal G}}^L$),  
are related through polarization-changing operators (\ref{operchan}) as 
follows:
\bea
\Delta_{\cp_2}^{(\pm)}(\th',\th)&=&\int\!\!\int_{\TG}{v(\tg')v(\tg)
\Delta_{\cp_2\cp_1}^{(\pm)}(\th',\tg')\Delta_{\cp_1}^{(\pm)}(\tg',\tg)
\Delta_{\cp_1\cp_2}^{(\pm)}(\tg,\th)}\nn\\
\Delta_{\cp_i\cp_j}^{(+)}(\th,\tg)&\equiv&\Delta_{\cp_i\cp_j}(\th,\tg)\,,
\;\;\;\Delta_{\cp_i\cp_j}^{(-)}(\th,\tg)\equiv\Delta_{\cp_i\cp_j}(\tg,\th)
\,.\label{polchanop}
\eea

To apply the GAQ formalism to $\tilde{G}^{(2)}$, it is appropriate to 
use a ``Fourier-like" parametrization, alternative to the 
field-like parametrization above [see (\ref{confpar})]. If we denote by
\be
\begin{array}{ll} a_n\equiv\langle n|f\rangle \,, & 
b_n\equiv\langle n^*|f\rangle \,,\\ a^*_n\equiv\langle f|n\rangle \,, & 
b^*_n\equiv\langle f|n^*\rangle \,,\end{array}\label{fupa}
\ee
\ni the Fourier coefficients of the ``particle" and the ``antiparticle", 
 a polarization subalgebra ${\cal P}^{(2)}$ 
for $\tilde{G}^{(2)}$ is always given by  
the corresponding left-invariant vector fields $\TXL_{a_n},\TXL_{b_n}$ and 
the whole Lie algebra $\tilde{{\cal G}}^L$ of $\TG$, which is the 
characteristic subalgebra 
${{\cal G}}_{\Theta^{(2)}}$ of 
the second-quantized theory (see next subsection). The operators of the 
theory are the right-invariant vector fields of $\tilde{G}^{(2)}$; 
in particular, the basic operators are: the annihilation operators of 
particles and anti-particles,  $\hat{a}_n\equiv\TXR_{a^*_n} ,\,
\hat{b}_n\equiv\TXR_{b^*_n}$, and the corresponding creation operators 
 $\hat{a}^{\dag}_n\equiv-\frac{1}{\kappa}\TXR_{a_n} ,\,\hat{b}^{\dag}_n\equiv 
-\frac{1}{\kappa}\TXR_{b_n}$. The operators corresponding to the subgroup 
$\TG$ [the second-quantized version $\TXRS_{\tg^j}$ of the first-quantized 
operators $\TXR_{\tg^j}$ in (\ref{txlr1})] are written in terms of the basic 
ones, since they are in the characteristic subalgebra 
${{\cal G}}_{\Theta^{(2)}}$ of the second-quantized theory.

The group $\TG$ plays a key role in picking out a preferred vacuum state and 
defining the notion of a ``particle", in the same way as the Poincar\'e 
group plays a central role in relativistic quantum theories defined on 
Minkowski space. 
In general, standard QFT in curved space suffers from the lack of a preferred 
definition of particles. The infinite-dimensional character of the symplectic 
solution manifold of a field system is responsible for the existence of an 
infinite number of unitarily inequivalent irreducible representations
of the Heisenberg-Weyl (H-W) relations and there 
is no criterion to select a preferred vacuum of the corresponding quantum 
field  (see, for example, \cite{Wald2,Birrell}). 
This situation is not present in the finite-dimensional case, 
according to the Stone-von Newman theorem \cite{Sharon,von}. 
In our language, the origen of this fact is related to the lack of a  
characteristic module for the H-W subgroup ${\tilde G}^{(2)}/\TG$ 
of ${\tilde G}^{(2)}$; i.e., for the infinite-dimensional H-W group, we 
can polarize the wave functions in arbitrary, non-equivalent directions. Thus, 
 so long as we can embed the (curved) 
space $M$ into a given group $\TG$, the existence of a characteristic module
-generated by ${\tilde{\cal G}}^L$- in the polarization subalgebra  helps  
us in picking out a preferred vacuum state. This vacuum state will be 
characterized by being annihilated by the right version of the 
polarization subalgabra dual to ${\cal P}^{(2)}$, i.e, it will be 
invariant under the action of 
$\TG\subset {\tilde G}^{(2)}$ and annihilated by 
the  right-invariant vector fields $\TXR_{a^*_n},\TXR_{b^*_n}$.

Other vacuum states might be selected as those 
states being 
invariant under  a subgroup $\TG_Q\subset \TG$ only, for example, 
the uniparametric subgroup of time evolution (see e.g. \cite{Bruce} for 
a discussion of vacuum states in de Sitter space). 
From our point of view, this 
situation would correspond to a breakdown of the symmetry and could be 
understood as a constrained theory of the original one. 
Indeed, let us  comment on the 
influence of the constraints 
in the first quantized theory at the second quantization level. 
Associated with
a constrained wave function satisfying (\ref{const}), there is a 
corresponding constrained quantum field subjected to the condition:
\be
\hbox{ad}_{\TXRS_{\tilde{t}}}\left(\TXR_{|f\rangle }\right)\equiv
\l[\TXRS_{\tilde{t}},\TXR_{|f\rangle }\r]=dD^{(\epsilon)}_T(\tilde{t})\TXR_{|f\rangle }\,,
\ee
\ni where  $\TXRS_{\tilde{t}}$ stands for the ``second-quantized version" of 
$\TXR_{\tilde{t}}$. It is straightforward to generalize the last condition to 
higher-order constraints: 
\bea
\TXR_1\TXR_2...\TXR_j|\psi\rangle &=&\epsilon |\psi\rangle 
\rightarrow\nn\\
\hbox{ad}_{\TXRS_1}\left(\hbox{ad}_{\TXRS_2}\left(...\hbox{ad}_{\TXRS_j}\left(
\TXR_{|f\rangle }\right)...\right)\right)&=&\epsilon\TXR_{|f\rangle }\,.
\label{goodfield}
\eea
\ni  The selection of a given Hilbert subspace 
${\cal H}^{(\epsilon)}(\TG)\subset {\cal H}(\TG)$ made of wave 
functions $\psi_c$ obeing a higher-order constraint 
$Q\psi_c=\epsilon\psi_c$, where 
$Q=\TXR_1\TXR_2...\TXR_j$ is some Casimir operator of $\TG_Q\subset \TG$,  
manifests, at second quantization level, as a new 
(broken) QFT. The vacuum for the new observables of this broken theory 
(the good operators in (\ref{goodfield})) does not have to coincide with 
the vacuum of the original theory, and the action of the rest of the operators 
(the bad operators) could make this new vacuum  radiate. This is precisely 
the problem we are involved, where $Q\equiv \TQR$ is the Casimir of the 
Poincar\'e subgroup inside the conformal group (see later in Sec. 4.2).

In general, constraints lead to gauge symmetries 
in the constrained theory and, also, 
the property for a subgroup $N\subset \TG$ 
of being gauge is heritable at the second-quantization level.

To conclude this subsection, it is important 
to note that the representation of $\TG$ on ${\cal M}(\TG)$ is reducible, but 
it is irreducible under $\TG$ together with the {\it charge conjugation} 
operation $a_n\leftrightarrow b_n$, which could be implemented on 
$\tilde{G}^{(2)}$. For simplicity, we have preferred to discard this 
transformation; 
however, a treatment including it, would be relevant as a revision of 
the CPT symmetry in quantum field theory. The Noether invariant associated 
with $\TXRS_\zeta$ is nothing other than the {\it total electric charge} 
(the total number of particles 
in the case of a real field 
$b_n\equiv a_n$) and its central character, inside the 
``dynamical" group $\TG$ of the first-quantized theory, now ensures its 
conservation under the action of the subgroup $\TG\subset\TG^{(2)}$. 
To account for non-abelian charges (iso-spin, color, etc), 
a non-abelian structure group 
$T\subset\TG$ is required.

\subsection{The case of the conformal group}

Let us now apply the GAQ formalism to the centrally extended 
group $\TG^{(2)}$ given through the group law in (\ref{group2}) for the case 
of $\TG=SO(2,2)$ and compact dilatation. We shall consider 
the case of a real field and we shall use a  ``Fourier" parametrization 
in terms of the coefficients $a_{n,\hn}$ rather than a ``field"  
parametrization in terms of $f_\cp(\tg)$. The explicit group law is:
\bea
\tg'' &=&\tg'*\tg \label{Fourier} \nn\\
{a_{m,\hm}}''&=&{a_{m,\hm}}'+\sum^\infty_{n,\hn=0}\A(\tg')a_{n,\hn} \nn\\
{a^*_{m,\hm}}''&=&{a^*_{m,\hm}}'+\sum^\infty_{n,\hn=0}\Ac(\tg')
a^*_{n,\hn} \label{acruz}\\
\vs''&=&\vs'\vs\exp{\f{\kappa}{2}\sum^\infty_{m,\hm=0}
\sum^\infty_{n,\hn=0}({a^*_{m,\hm}}'
\A(\tg')a_{n,\hn}-{a_{m,\hm}}'\Ac(\tg')a^*_{n,\hn})}\,.\nn
\eea

The left- and right-invariant vector fields (we denote $\p_{m,\hm}
\equiv\f{\p}{\p a_{m,\hm}},\;
\p^*_{m,\hm}\equiv\f{\p}{\p a^*_{m,\hm}}$) are:
\bea
\TXL_\vs&=&\TXR_\vs=\vs\f{\p}{\p \vs}\nn \\
\TXL_{a_{n,\hn}}&=&\sum^\infty_{m,\hm=0}\A(\tg){\p_{m,\hm}}+\f{\kappa}{2}
\sum^\infty_{m,\hm=0}\A(\tg)a^*_{m,\hm}\TXL_\vs\nn \\
\TXL_{a^*_{n,\hn}}&=&\sum^\infty_{m,\hm=0}\Ac(\tg){\p^*_{m,\hm}}
-\f{\kappa}{2}
\sum^\infty_{m,\hm=0}\Ac(\tg)a_{m,\hm}\TXL_\vs\nn \\
\TDLS &=&\TDL,\,\,\,\TMLS =\TML,\,\,\,\TPOLS =\TPOL,\nn\\
\TPLS &=&\TPL,\,\,\,\TKOLS =\TKOL,\,\,\,\TKLS =\TKL,\,\,\,
\TXLS_\zeta=\TXL_\zeta\nn\\
\TXR_{a_{n,\hn}}&=&{\p_{n,\hn}}-\f{\kappa}{2}a^*_{n,\hn}\TXL_\vs\nn \\
\TXR_{a^*_{n,\hn}}&=&{\p^*_{n,\hn}}+\f{\kappa}{2}a_{n,\hn}\TXL_\vs\nn \\
\TDRS &=&\TDR-\sum^\infty_{m,\hm=0}(m+\hm)(a_{m,\hm}{\p_{m,\hm}}-
a^*_{m,\hm}{\p^*_{m,\hm}})\nn\\
\TMRS &=&\TMRS+\sum^\infty_{m,\hm=0}(m-\hm)(a_{m,\hm}{\p_{m,\hm}}-
a^*_{m,\hm}{\p^*_{m,\hm}})\nn\\
\TPORS&=&\TPOR+\sum^\infty_{m,\hm=0}\left(\sqrt{(m+1)(2N+m)}
(a_{m+1,\hm}\p_{m,\hm}-a^*_{m,\hm}\p^*_{m+1,\hm})\right.\nn\\
&+&\left.\sqrt{(\hm+1)(2N+\hm)}
(a_{m,\hm+1}\p_{m,\hm}-a^*_{m,\hm}\p^*_{m,\hm+1})\right)\nn\\
\TPRS &=&\TPR-\sum^\infty_{m,\hm=0}\left(\sqrt{(m+1)(2N+m)}
(a_{m+1,\hm}\p_{m,\hm}-a^*_{m,\hm}\p^*_{m+1,\hm})\right.\nn\\
&-&\left.\sqrt{(\hm+1)(2N+\hm)}
(a_{m,\hm+1}\p_{m,\hm}-a^*_{m,\hm}\p^*_{m,\hm+1})\right)\nn\\
\TKORS &=&\TKOR+\sum^\infty_{m,\hm=0}\left(\sqrt{(m+1)(2N+m)}
(a_{m,\hm}\p_{m+1,\hm}-a^*_{m+1,\hm}\p^*_{m,\hm})\right.\nn\\
&+&\left.\sqrt{(\hm+1)(2N+\hm)}
(a_{m,\hm}\p_{m,\hm+1}-a^*_{m,\hm+1}\p^*_{m,\hm})\right)\nn\\
\TKRS &=&\TKR+\sum^\infty_{m,\hm=0}\left(\sqrt{(m+1)(2N+m)}
(a_{m,\hm}\p_{m+1,\hm}-a^*_{m+1,\hm}\p^*_{m,\hm})\right.\nn\\
&-&\left.\sqrt{(\hm+1)(2N+\hm)}
(a_{m,\hm}\p_{m,\hm+1}-a^*_{m,\hm+1}\p^*_{m,\hm})\right)\nn\\
\TXRS_\zeta&=&\TXR_\zeta-\sum^\infty_{m,\hm=0}(a_{m,\hm}{\p_{m,\hm}}-
a^*_{m,\hm}{\p^*_{m,\hm}}).
\eea
\ni  The non-trivial commutators between 
those vector fields are:
\bea
\l[\TXL_{a_{n,\hn}},\TXL_{a^*_{m,\hm}}\r]&=&-\kappa\delta_{nm}
\delta_{\hn\hm}\TXL_{\vs}\nn\\
\l[\TDLS,\TXL_{a_{n,\hn}}\r]&=&-(n+\hn)\TXL_{a_{n,\hn}}\nn\\
\l[\TMLS,\TXL_{a_{n,\hn}}\r]&=&(n-\hn)\TXL_{a_{n,\hn}}\nn\\
\l[\TPOLS,\TXL_{a_{n,\hn}}\r]&=&\sqrt{n(2N+n-1)}\TXL_{a_{n-1,\hn}}
+\sqrt{\hn(2N+\hn-1)}\TXL_{a_{n,\hn-1}}\nn\\
\l[\TPLS,\TXL_{a_{n,\hn}}\r]&=&-\sqrt{n(2N+n-1)}\TXL_{a_{n-1,\hn}}
+\sqrt{\hn(2N+\hn-1)}\TXL_{a_{n,\hn-1}}\nn\\
\l[\TKOLS,\TXL_{a_{n,\hn}}\r]&=&\sqrt{(n+1)(2N+n)}\TXL_{a_{n+1,\hn}}
+\sqrt{(\hn+1)(2N+\hn)}\TXL_{a_{n,\hn+1}}\nn\\
\l[\TKLS,\TXL_{a_{n,\hn}}\r]&=&\sqrt{(n+1)(2N+n)}\TXL_{a_{n+1,\hn}}
-\sqrt{(\hn+1)(2N+\hn)}\TXL_{a_{n,\hn+1}}\nn\\
\l[\TXLS_\zeta,\TXL_{a_{n,\hn}}\r]&=&\TXL_{a_{n,\hn}}\nn\\
\l[\TDLS,\TXL_{a^*_{n,\hn}}\r]&=&(n+\hn)\TXL_{a^*_{n,\hn}}\nn\\
\l[\TMLS,\TXL_{a^*_{n,\hn}}\r]&=&-(n-\hn)\TXL_{a^*_{n,\hn}}\nn\\
\l[\TPOLS,\TXL_{a^*_{n,\hn}}\r]&=&-\sqrt{(n+1)(2N+n)}\TXL_{a^*_{n+1,\hn}}
-\sqrt{(\hn+1)(2N+\hn)}\TXL_{a^*_{n,\hn+1}}\nn\\
\l[\TPLS,\TXL_{a^*_{n,\hn}}\r]&=&\sqrt{(n+1)(2N+n)}\TXL_{a^*_{n+1,\hn}}
-\sqrt{(\hn+1)(2N+\hn)}\TXL_{a^*_{n,\hn+1}} \nn\\
\l[\TKOLS,\TXL_{a^*_{n,\hn}}\r]&=&-\sqrt{n(2N+n-1)}\TXL_{a^*_{n-1,\hn}}
-\sqrt{\hn(2N+\hn-1)}\TXL_{a^*_{n,\hn-1}}\nn\\
\l[\TKLS,\TXL_{a^*_{n,\hn}}\r]&=&-\sqrt{n(2N+n-1)}\TXL_{a^*_{n-1,\hn}}
+\sqrt{\hn(2N+\hn-1)}\TXL_{a^*_{n,\hn-1}}\nn\\
\l[\TXLS_\zeta,\TXL_{a^*_{n,\hn}}\r]&=&-\TXL_{a^*_{n,\hn}}\,,\label{2ndconf}
\eea
\ni where we have omitted the commutators corresponding to the 
extended conformal subgroup, which have the same form as in 
(\ref{conforme}), except for the two commutators in 
(\ref{confcomp}).

The quantization 1-form  and the characteristic module are:
\bea
\Theta^{(2)}&=&\f{i\kappa}{2}\sum^\infty_{n,\hn=0}
(a_{n,\hn}da^*_{n,\hn}-a^*_{n,\hn}da_{n,\hn})-i\vs^{-1}d\vs \nn\\
{{\cal G}}_{\Theta^{(2)}}&=&<\TDLS,\TMLS,\TPOLS,\TPLS,\TKOLS,
\TKLS,\TXLS_\zeta>\;. \label{2form}
\eea

A full polarization subalgebra is:
\be
{\cal P}^{(2)}=<{{\cal G}}_{\Theta^{(2)}},\TXL_{a_{n,\hn}}>, 
\;\;\forall n,\hn\geq 0
\ee
\ni and the polarized $U(1)$-functions have the form:
\be
\Psi[a,a^*,\tg,\vs]=\vs\exp\left\{-\f{\kappa}{2}
\sum^\infty_{n,\hn=0}a^*_{n,\hn}a_{n,\hn}
\right\}\Phi[a^*]\equiv \vs\Omega\Phi[a^*]\,,\label{2wavepol}
\ee
\ni where $\Omega$ is the vacuum of the second quantized 
theory and $\Phi$ is an arbitrary power series in its argument.

The actions of the right-invariant vector fields 
(operators in the second-quantized theory) on
polarized wave functions in (\ref{2wavepol}) have the explicit form: 
\bea
\TXR_{a_{n,\hn}}\Psi&=&\vs\Omega\cdot(-\kappa a^*_{n,\hn})\Phi\equiv 
\vs\Omega\cdot(-\kappa {\hat{a}}^{\dag}_{n,\hn})\Phi\nn\\
\TXR_{a^*_{n,\hn}}\Psi&=&\vs\Omega\cdot({\p^*_{n,\hn}})\Phi\equiv 
\vs\Omega\cdot({\hat{a}}_{n,\hn})\Phi\nn\\
\TDRS\Psi&=&\vs\Omega\cdot\left(\sum^\infty_{n,\hn=0}(n+\hn)
\ac_{n,\hn}\a_{n,\hn}\right)\Phi\nn\\
\TMRS\Psi&=&\vs\Omega\cdot\left(-\sum^\infty_{n,\hn=0}(n-\hn)
\ac_{n,\hn}\a_{n,\hn}\right)\Phi\nn\\
\TPORS\Psi&=&\vs\Omega\cdot\left(-\sum^\infty_{n,\hn=0}
\sqrt{(n+1)(2N+n)}\ac_{n,\hn}\a_{n+1,\hn}+
\sqrt{(\hn+1)(2N+\hn)}\ac_{n,\hn}\a_{n,\hn+1}\right)\Phi\nn\\
\TPRS\Psi&=&\vs\Omega\cdot\left(\sum^\infty_{n,\hn=0}
\sqrt{(n+1)(2N+n)}\ac_{n,\hn}\a_{n+1,\hn}-
\sqrt{(\hn+1)(2N+\hn)}\ac_{n,\hn}\a_{n,\hn+1}\right)\Phi\nn\\
\TKORS\Psi&=&\vs\Omega\cdot\left(-\sum^\infty_{n,\hn=0}
\sqrt{(n+1)(2N+n)}\ac_{n+1,\hn}\a_{n,\hn}+
\sqrt{(\hn+1)(2N+\hn)}\ac_{n,\hn+1}\a_{n,\hn}\right)\Phi\nn\\
\TKRS\Psi&=&\vs\Omega\cdot\left(-\sum^\infty_{n,\hn=0}
\sqrt{(n+1)(2N+n)}\ac_{n+1,\hn}\a_{n,\hn}-
\sqrt{(\hn+1)(2N+\hn)}\ac_{n,\hn+1}\a_{n,\hn}\right)\Phi\nn\\
\TXRS_\zeta\Psi&=&\vs\Omega\cdot\left(\sum^\infty_{n,\hn=0}\ac_{n,\hn}
\a_{n,\hn}\right)\Phi
\eea
\ni where $\a_{n,\hn}$ and $\ac_{n,\hn}$ are interpreted as annihilation 
and creation 
operators of {\it modes} $|n;\hn\rangle   $, $\TDRS$ is 
attached to the total energy (remember that the dilatation 
parameter plays the role of a proper time), and $\TXLS_\zeta$ 
corresponds with the 
number operator. It should be mentioned that all those 
quantities appear, in a natural way, {\it normally ordered}; this is one 
of the advantages of this method of quantization: normal order does not have 
to be imposed by hand but, rather, it is implicitly inside the 
formalism itself.

We can think of the Hilbert space as composed of modes: 
\begin{enumerate}
\item pure non-bar $|n_1,n_2,...;0\rangle   $ \,,
\item pure bar $|0;\hn_1,\hn_2,...\rangle   $ \,, 
\item mixed $|n_1,n_2,...;\hn_1,\hn_2,...\rangle   $\,. 
\end{enumerate}

\subsection{Breaking down to the Weyl subgroup. Vacuum radiation}

In this subsection, we investigate the effect of SCT on a Weyl vacuum, i.e. 
a vacuum of the massless QFT obtained after constraining  the 
conformal quantum field theory 
developed in the last subsection.

The field degrees of freedom of the massless field are obtained by 
translating the condition  (\ref{kglicke}) to the second quantization 
level, according to the general procedure (\ref{goodfield}) that is, by 
imposing 
\bea
 \l[\TPORS+\TPRS,\l[\TPORS-\TPRS, \TXR_{a_{n,\hn}}\r]\r]&=&\nn\\ & & \nn\\
-4\sqrt{n(2N+n-1)}\sqrt{\hn(2N+\hn-1)}\TXL_{a_{n-1,\hn-1}}&=&0\,,
\eea
\ni which selects the pure non-bar and pure bar operators, i.e, 
$\ac_{n,0}=-\frac{1}{\kappa}\TXR_{a_{n,0}}$ and 
$\ac_{0,\hn}=-\frac{1}{\kappa}\TXR_{a_{0,\hn}}$. These operators, 
together with the Weyl generators (good operators of the first-quantized 
theory) close a   Lie subalgebra
\be
{\cal G}^{(2)}_c=<\TDRS,\TMRS,\TPORS,\TPRS, \TXLS_\zeta,\ac_{n,0},\ac_{0,\hn}>
\ee
\ni of the original Lie algebra of the conformal quantum field. The vacuum 
of this constrained theory does not have to coincide with the conformal 
vacuum $|0\rangle =|n=0;\hn=0\rangle $. In fact, any conformal 
state made up of an arbitrary content of zero modes
\be
|W\{\sigma\}\rangle   \equiv\sum^\infty_{q=0}\sigma_q(\ac_{0,0})^q
|0\rangle   \label{Weylvac}
\ee
\ni behaves as a vacuum from the point of view of a {\it Weyl observer},  
that is, 
it is annihilated by the Weyl generators and the destruction  operators 
$\a_{n,0}$ and $\a_{0,\hn}$, for all $n,\hn\in N-\{0\}$. Note that, since the 
operator $\ac_{0,0}$ is central in ${\cal G}^{(2)}_c$ (it commutes with 
all the others), it would be too restrictive to require 
(\ref{Weylvac}) being nullified by $\a_{0,0}$; the only solution 
would be the conformal vacuum $|0\rangle$. It is then natural to 
demand that $\a_{0,0}$ behave as a multiple $\vartheta$ of 
the identity that is, it has to 
leave the Weyl vacuum stable 
\be
\a_{0,0}|W\{\sigma\}\rangle =\vartheta 
|W\{\sigma\}\rangle\Rightarrow 
\sigma^{(0)}_q=\frac{\vartheta^q}{q!}\sigma_0\,,\label{stable}
\ee
\ni condition which, after normalizing, determines the Weyl vacuum 
up to a complex parameter $\vartheta$
\be
\langle W\{\sigma^{(0)}\}|W\{\sigma^{(0)}\}\rangle   =1\Rightarrow 
|\sigma_0|=e^{-\frac{1}{2}|\vartheta|^2}\,.
\ee
\ni Thus, we have find a set of  Weyl vacua (coherent states of 
the conformal  quantum field, made of zero modes)
\be
 |0\rangle_\vartheta \equiv e^{-\frac{1}{2}|\vartheta|^2}
e^{\vartheta\ac_{0,0}}|0\rangle   \,, \label{theta}
\ee
\ni labeled by $\vartheta$ [the existence of a degenerate 
ground state resembles the 
``$\theta$-vacuum'' phenomenon in Yang-Mills field theories \cite{AEP} and, 
in general,  it is present whenever we deal with non-simply connected 
phase spaces and constrained theories \cite{FracHall}]. 
As the final result is independent 
of $\vartheta$, from now on we shall implicitly 
choose $\vartheta=1$, for the sake of 
simplicity. An orthonormal basis  
for the Hilbert space of the constrained theory can be obtained by 
taking the orbit 
through the vacuum (\ref{theta}) of the creation operators as follows:
\be
|m(n_1),...,m(n_q),m(\hn_1),...,m(\hn_j)\rangle_\vartheta\equiv
\frac{(\ac_{n_1,0})^{m(n_1)}...(\ac_{n_q,0})^{m(n_q)}
(\ac_{0,\hn_1})^{m(\hn_1)}...(\ac_{0,\hn_j})^{m(\hn_j)}}{(m(n_1)!...m(n_k)!
m(\hn_1)!...m(\hn_j)!)^{1/2}}|0\rangle_\vartheta   \,.\label{estados}
\ee
\ni We can make a comparison with the standard case of a massless field 
in 1+1 dimensional Minkowski space-time and relate the non-bar and bar modes 
to the left-hand and right-hand moving scalar photons, respectively. 
Let us introduce dimensions through the Planck constant $h$ and the 
frequency mode $\nu$, so that the total energy is given by 
\be
\hat{E}\equiv h\nu\left(\TDRS+2N\TXLS_\zeta\right)\equiv h\nu D^{R(2)}\,;
\ee
\ni the last redefinition of the dilatation generator is intended to 
render the commutation relations (\ref{confcomp}) to the usual ones 
(\ref{conforme}) by 
destroying the pseudo-extension (\ref{cdp}). The expected value of the 
energy in the general state (\ref{estados}) is 
\be
\langle \hat{E}\rangle=h\nu(\sum^q_{l=1}m(n_l)n_l+\sum^j_{l=1}m(\hn_l)\hn_l
+2N)\,,
\ee
\ni where $E_0\equiv 2N h\nu$ represents the {\it zero point energy}, 
i.e. the expected value of the energy in the Weyl vacuum.
Zero modes represent {\it virtual} particles (they have no energy and 
cannot be detected by a Weyl observer) 
and can be spontaneously created from the Weyl vacuum, as can be deduced 
from the condition (\ref{stable}). 

It is natural to think that zero modes will 
play an important role in the radiation of a Weyl vacuum, as they will be made 
{\it real} by  acceleration. In fact, let us 
show how a finite special conformal 
transformation, generated by $A^{(2)}_+\equiv\frac{1}{2}(\TKOLS+\TKLS)$, 
applied to a Weyl vacuum 
gives rise to a ``thermal bath" of  no-bar modes (left-hand moving scalar
 photons), 
whereas the combination $A^{(2)}_-\equiv\frac{1}{2}(\TKOLS-\TKLS)$ gives 
rise to a 
``thermal bath" 
of bar modes (right-hand moving scalar photons). The finite action of 
$A^{(2)}_+$, with 
parameter $\alpha$ (the corresponding acceleration is 
$a\equiv -(2\pi)^2\frac{c\nu}{\log|\alpha|^2}$, where $c$ is the speed of 
light), 
on the Fourier parameter,
\bea
a^*_{0,0}\rightarrow {a^*_{0,0}}'&=& \sum^\infty_{n=0}(-1)^n
\sqrt{\frac{C^{(N)}_0}{C^{(N)}_n}} a^*_{n,0}\alpha^n = 
\sum^\infty_{n=0}r_n a^*_{n,0}\alpha^n \nn\\
r_n &\equiv& (-1)^n
\sqrt{\frac{C^{(N)}_0}{C^{(N)}_n}}
\eea 
\ni (according to the general expression in the third line of 
(\ref{acruz}) and the last equality  in Eq. (\ref{fouritrans})), leads 
to the following transformation on the Weyl vacuum:
\be
|0\rangle_\vartheta \rightarrow |\Psi(\alpha)\rangle_\vartheta\equiv
 e^{-\frac{1}{2}}e^{\ac_{0,0}{}'}|0\rangle=  \sum^\infty_{q=0}
\alpha^q\sum_{\begin{array}{c}
m_1,...,m_q :\\ \sum^q_{n=0}n m_n=q\end{array}}
\prod^q_{n=0}\frac{r_n^{m_n}}{m_n!}
\prod^q_{n=0}(\ac_{n,0})^{m_n}|0\rangle_\vartheta \,,\label{bath}
\ee
\ni where $m_0=0$ and we have used the general identity
\bea
\left(\sum^\infty_{n=0}\gamma_n \c^n\right)^l&=&\sum^\infty_{q=0} 
\delta_q \c^q\nn\\
\delta_0&=&\gamma^l_0\nn\\
\delta_q&=&\frac{1}{q\gamma_0}\sum^q_{s=1}(sm-q+s)\gamma_s\delta_{q-s}\,.
\eea
\ni The relative probability of observing a state with total energy 
$E_q=h\nu q+E_0$ in a Weyl vacuum from an accelerated frame (i.e. in 
$|\Psi(\alpha)\rangle_\vartheta$) is  
\bea
P_q&=&\Lambda(E_q)(|\alpha|^2)^q \nn\\
\Lambda(E_q)&\equiv &\sum_{\begin{array}{c}
m_1,...,m_q :\\ \sum^q_{n=0}n m_n=q\end{array}}
\prod^q_{n=0}\frac{r_n^{2m_n}}{m_n!}\,.
\eea
\ni We can associate a thermal bath with this distribution function by 
noticing that $\Lambda(E_q)$ represents a relative weight 
proportional to the number of states with energy $E_q$, and the factor 
$(|\alpha|^2)^q$ fits this weight properly to a temperature as
\be
(|\alpha|^2)^q=e^{q\log|\alpha|^2}=e^{-\frac{E_q-E_0}{kT}}\,,\,\,\,\;
\hbox{where}\,\,\,\,\; T\equiv-\frac{h\nu}{k\log|\alpha|^2}=\frac{\hbar  a}
{2\pi ck} 
\label{temper}
\ee
\ni is the temperature associated with a given acceleration $a$, and 
$k$ is the Boltzmann's 
constant. This simple, but profound, relation between temperature and 
acceleration  was first considered by  Unruh \cite{Unruh}.
The balance between the ``multiplicity factor'' $\Lambda(E_q)$ (an increasing 
function of the energy) and the temperature factor (\ref{temper}) 
(a decreasing function of the energy) is favorable (maximum) for a 
given system of this canonical ensemble, the energy of which is a 
repersentative value of the mean energy. In fact, this mean energy 
can be calculated exactly as the expected value of the 
energy operator $\hat{E}$ in the state $|\Psi(\alpha)\rangle_\vartheta$. 
To this end, let us perform some intermediate calculations. The norm of 
this accelerated 
vacuum is
\be
\hbox{Nor}[\Psi(\alpha)]\equiv {}_\vartheta\langle\Psi(\alpha)|
\Psi(\alpha)\rangle_\vartheta=
\exp\left(-1+\sum^\infty_{n=0}r^2_n
|\alpha|^{2n}\right)=\exp\left((1-|\alpha|^2)^{-2N}-1\right)\,.
\ee
\ni The probability $P_n(m)$ of observing $m$ particles with energy 
$E_n$ coincides 
with the expected value of the projector $\hat{P}_n(m)$ on the state 
$|m(n)\rangle_\vartheta$, i.e.:
\be
P_n(m)\equiv\frac{{}_\vartheta\langle\Psi(\alpha)|\hat{P}_n(m)
|\Psi(\alpha)\rangle_\vartheta }{\hbox{Nor}[\Psi(\alpha)]}
=\frac{1}{\hbox{Nor}[\Psi(\alpha)]}\,
\frac{(r^{2}_n |\alpha|^{2n})^{m}}{m!}\,;
\ee
\ni it  can be seen that the closure relation $\sum^\infty_{n,m=0}P_n(m)=1$ 
is in fact verified. The 
mean number ${\cal N}_n$ of left-hand moving scalar photons with energy $E_n$ 
corresponds with the expected value of the 
number operator $\hat{\cal{N}}_n\equiv\ac_{n,0}\a_{n,0}\,$, i.e. :
\be
{\cal N}_n = \sum^\infty_{m=0}mP_n(m)=\frac{1}{\hbox{Nor}[\Psi(\alpha)]}
\,\, r^{2}_n |\alpha|^{2n}\exp\left(r^{2}_n |\alpha|^{2n}\right)\,.
\ee
\ni With this information at hand, we can calculate the expected value of 
the total energy as:
\be
E[\Psi(\alpha)]\equiv E_0+h\nu\sum^\infty_{n=1} n{\cal N}_n=
E_0+\frac{2Nh\nu|\alpha|^2}{(1-|\alpha|^2)^{2N+1}}\,.
\ee
\ni If we subtract the zero-point energy, normalize by a $2N$ factor 
(this normalization can be seen as a reparametrization of the proper time) 
and make use of the relation $|\alpha|^2=e^{-\frac{h\nu}{kT}}$ in 
(\ref{temper}),  we obtain a 
more ``familiar'' expression for the {\it mean energy per mode} 
\be
\epsilon_N(\nu,T)=\frac{h\nu e^{-\frac{h\nu}{kT}}}{(1-
e^{-\frac{h\nu}{kT}})^{2N+1}}\,,
\ee
\ni  the value  $N=0$ corresponding with the well known case of the 
{\it Bose-Einstein statistic}. Note that this particular 
value of $N$ can be reached 
only as a limiting process formulated on the universal covering of 
$SU(1,1)\otimes SU(1,1)$ or, equivalently, by uncompactifying the 
proper time ($U(1)\rightarrow \Re$).

Let us compare 
 the spectral distribution of the radiation of the Weyl vacuum for 
different values of $N$, with the well 
known case of the {\it black body radiation} (Planck's spectrum). For this 
purpose, we have to multiply the mean energy per mode by the number 
of states with frequency 
$\nu$ which, in $d$ dimensions, would be proportional to $\nu^{d-1}$. 
If we denote this product by 
$u(x,N)\equiv u_0\,x^{d-1}\epsilon_N(x,T_0)$ with 
$x\equiv\frac{h\nu}{kT_0}$  ($u_0$ is a constant, for a 
fixed temperature $T_0$, with dimensions of energy per unit of 
volume),  Figure 1 shows the departure from 
the Planckian spectrum ($N=0$) for 
four diferent values $N=\frac{1}{2},\,\frac{3}{4}\,,1,\,1.1\,$ in 
the realistic dimension $d=3$. Note that the value of $N\equiv N_c=
\frac{d-1}{2}$ corresponds to a {\it critical} situation: over this value 
the theory exhibits an ``infrared catastrophe''.

\section{Other representations: some comments}

We have shown that it is impossible to establish conformally 
invariant evolution 
equations in a (even compactified) Minkowski space, 
not only for massive fields 
but, also for massless {\it quantum} fields. 

If we wish the whole conformal group to be an exact symmetry of  physical 
laws (at least, at very high energies), then we should reconsider the 
convenience of the Minkowski space as the frame for describing quantum 
physical phenomena. In fact, there exists a wider consistent quantum dynamics 
in which the conformal invariance is exact. The price to be paid is the 
introduction of an extra dimension, thus increasing by one the number 
of space-time parameters. The physical interpretation
of this new parameter remains obscure but, 
interpretations in terms of a ``unit of measurement''
\`a la Weyl \cite{invmass} and/or a ``variable mass'' 
interpretation have already been treated in the
literature, even at the non-relativistic (Galilean) level \cite{VarMass}. 
 
In the present scheme, this kind  of representation can be obtained through 
a higher-order polarization which, as we are going to see, carries an index 
$m_0$ eventually interpreted as a conformally invariant mass (see 
\cite{invmass}). In fact, let us use  the following 
couples of generators:
\bea
\TPIOL\equiv \f{\sqrt{2}}{2}(\TPOL+\TKOL) &;&
\TXOL\equiv \f{\sqrt{2}}{2}(\TPOL-\TKOL) \nn \\
&\hbox{and} & \\
\TPIL\equiv \f{\sqrt{2}}{2}(\TPL+\TKL) &;&  
\TX1L\equiv \f{\sqrt{2}}{2}(\TPL-\TKL)\nn
\eea

\ni as conjugate variables. If one tries to introduce  the set 
$\{\TXOL,\TX1L\}$  in the polarization, then 
we face the problem that the characteristic module generated by the operators
$\TDL,\TML$ is too large, since 
$\l[\tilde{X}_\nu^L, \tilde{D}^L\r]=-\tilde{\Pi}_\nu^L$; in fact, 
only $\TML$ possesses a compatible set of commutation
relations. As we have already pointed 
out, the dilatation could be introduced at the price 
of being a higher-order operator 
(something similar occurs with the time operator
in the free particle case as long as we stay in position representation). 
More precisely, with this higher-order 
polarization, one can reduce the representation
as follows:

\be
{\cal P}=<\TML,{\TXL}_0,{\TXL}_1,\tilde{C}^L>\,,
\ee
\ni where $\tilde{C}^L$ is just a Casimir operator of the extended 
conformal group [we can always add  an arbitrary central 
term to $\tilde{C}^L$]. For example, in the compact dilatation case
\be
\tilde{C}^L=({\TML})^2+({\TDL}+2N\TXL_{\zeta})^2-
({\tilde{\Pi}^L})^2+({\tilde{X}^L})^2\,.
\ee
\ni   Polarized wave functions evolve according to a 
Klein-Gordon-like equation 
\be
({\tilde{\Pi}^L})^2\psi=({\TDL}+2N)^2\psi\, ,\label{ultima}
\ee 
\ni which can be interpreted as the motion equation of a scalar field 
with variable square mass $m^2=({\TDL}+2N)^2=(D^L)^2$. 
The value of the Casimir on polarized wave functions is  $\tilde{C}^R\psi=
N(N-1)\psi\equiv m_0\psi$, which justify the denomination of $m_0$ 
as a conformally invariant mass [it proves to be quantized for this case, 
the  reason being related to the compact character of the proper time 
(dilatation)]. The allowed value of $N$, $N=1$ thus corresponds 
with null conformal mass. The precise connection between $N$ and 
the curvature of some homogeneous subspaces (let us say, the Anti-de 
Sitter universe  in $2+1$ dimensions)  inside the conformal 
group is being investigated \cite{prepara}.

Note that the Cauchy hyphersurfaces of Eq. (\ref{ultima}) have dimension 2, and
the physical interpretation of the extra dimension remains unclear. 
Two different approaches can be taken which 
could be consistent with the physical meaning
 of the conformal group. One is related to 
the Weyl idea of different lengths 
in different points of space time \cite{Weyl}. 
The ``rule" for measuring distances changes at 
different positions. In Quantum Mechanics, this 
implies that wave functions measuring 
probability densities do have different 
integration measures as functions of space-time.
This change in the measure of integration needs 
to be related to the extra parameter 
appearing in our Group Approach to Quantization 
of the full conformal group.
The other interpretation --not necessarily unrelated 
to the previous one-- could be attached to 
the variable character of mass. Even at the level 
of one particle ordinary conformal quantum 
mechanics, the inescapable consequence of a variable 
mass  appearing in the formalism was 
already observed several years ago by Niederer 
\cite{VarMass}. It indeed would not be  a surprise should this 
fact  also have some consequences in the full quantization. 
Neither interpretation, however, 
is without controversy, as  
emphasized previously by Rohrlich \cite{Horror}.
At any rate, we have considered here a more 
satisfactory point of view by examining the dynamical
breaking of the conformal group  down to the Weyl subgroup
in the framework of the Group Approach to 
Quantization.

\section*{Acknowledgements}

M. Calixto thanks the Spanish MEC for a FPI grant. M.C. is 
also grateful to A.P. Balachandran for his hospitality at the Department of 
Physics of the Syracuse University, where some of the work on the manuscript 
was carried out.

%\vskip 0.5 cm

%\newpage

\newpage
\begin{figure}[hbtw]
\epsfxsize=15.0cm
\centerline{\epsfbox{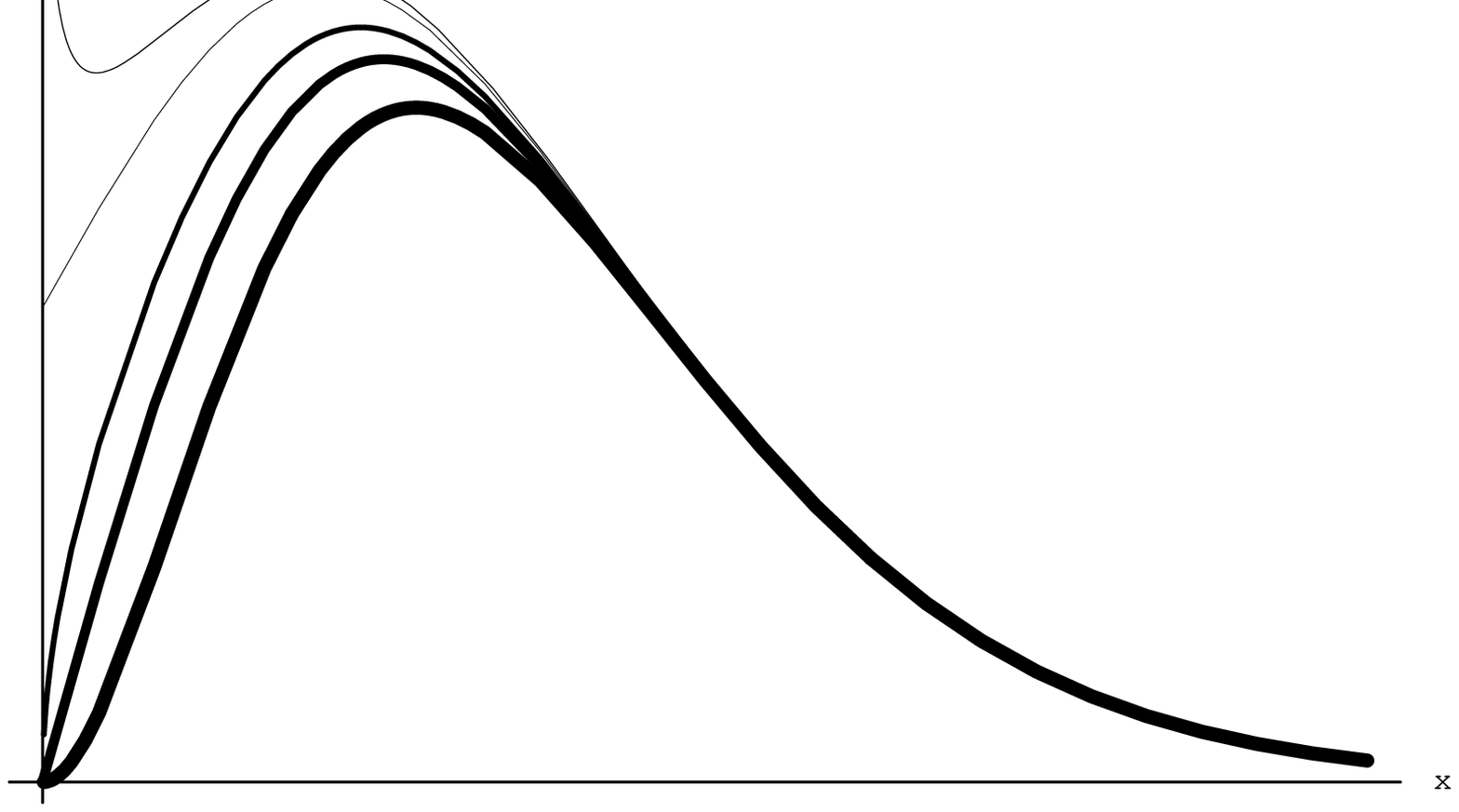}}
\vspace{-13.6cm}
\caption{Departure from the Planck's spectrum (tickest line) for increasing 
values of $N$ (decreasing tickness).}
\end{figure}

\end{document}